\documentclass[twocolumn,10pts,nofootinbib,superscriptaddress,preprintnumbers]{revtex4-2}
\usepackage[utf8]{inputenc}
\usepackage[normalem]{ulem}

\def\mySections#1{{\bf #1.} } 
\usepackage{hyperref}
\usepackage{braket}  
\usepackage[dvips]{graphicx}
\usepackage{hhline}
\usepackage{epsfig,amsmath,amssymb,verbatim,mathrsfs,array,layout,textcomp,amssymb,latexsym,slashed,graphicx,booktabs,color,mathtools}
\usepackage{bm}

\newcommand{\beq}{\begin{equation}}% can be used as {equation} or {eqnarray}
\newcommand{\eeq}{\end{equation}}

\def\beqa{\begin{eqnarray}}
\def\eeqa{\end{eqnarray}}
\def\bea{\begin{eqnarray}}
\def\eea{\end{eqnarray}}

\newcommand{\bv}{\left(\begin{array}{c}}
\newcommand{\ev}{\end{array}\right)}
\newcommand{\bmtwo}{\left(\begin{array}{cc}}
\newcommand{\bmthree}{\left(\begin{array}{ccc}}
\newcommand{\emn}{\end{array}\right)}
\newcommand{\bmtwoc}{\left\{\begin{array}{cc}}
\newcommand{\bmthreec}{\left\{\begin{array}{ccc}}
\newcommand{\emnc}{\end{array}\right\}}
\newcommand{\ba}{\begin{array}}
\newcommand{\ea}{\end{array}}
\newcommand{\be}{\begin{equation}}
\newcommand{\ee}{\end{equation}}

\let\originalleft\left
\let\originalright\right
\renewcommand{\left}{\mathopen{}\mathclose\bgroup\originalleft}
\renewcommand{\right}{\aftergroup\egroup\originalright}

\newcommand{\GeV}{\text{ GeV}}

\definecolor{readableRTD}{rgb}{0.7,0.1,0.2}

\newcommand{\tn}[1]{\textnormal{#1}}

\DeclareMathOperator{\Tr}{Tr}

\definecolor{readableMG}{rgb}{0.0,0,0.5}

\def\lsim{\mathrel{\rlap{\lower4pt\hbox{\hskip1pt$\sim$}}
     \raise1pt\hbox{$<$}}}         %less than or approx. symbol
\def\gsim{\mathrel{\rlap{\lower4pt\hbox{\hskip1pt$\sim$}}
     \raise1pt\hbox{$>$}}}         %greater than or approx. symbol

\definecolor{bluDT}{cmyk}{1,0.5,0,0.3}

\definecolor{darkblue}{rgb}{0.2,0.2,0.9}
\definecolor{colorRTD}{rgb}{.7,.2,.2}

\usepackage{soul}

\makeatletter
\renewcommand{\p@subsection}{}
\renewcommand{\p@subsubsection}{}
\makeatother

\begin{document}

\title{A Multiverse Outside of the Swampland} 

\author{Raffaele Tito D'Agnolo}
\affiliation{Université Paris-Saclay, CNRS, CEA, Institut de Physique Théorique, 91191, Gif-sur-Yvette, France}
\affiliation{Laboratoire de Physique de l’École Normale Supérieure, ENS, Université PSL, CNRS, Sorbonne
Université, Université Paris Cité, 75005, Paris, France}

\author{Paolo Mangini}
\affiliation{Dipartimento di Fisica, Sapienza Università di Roma, Piazzale Aldo Moro 5, 00185, Roma, Italy}

\author{Gabriele Rigo}
\affiliation{Université Paris-Saclay, CNRS, CEA, Institut de Physique Théorique, 91191, Gif-sur-Yvette, France}

\author{Lian-Tao Wang}
\affiliation{Department of Physics, University of Chicago, Chicago, IL 60637, USA}
\affiliation{Enrico Fermi Institute, University of Chicago, Chicago, IL 60637, USA}
\affiliation{Kavli Institute for Cosmological Physics, University of Chicago, Chicago, IL 60637, USA}

\begin{abstract}
A Multiverse can arise from landscapes without de Sitter minima. It can be populated during a period of eternal inflation without trans-Planckian field excursions and without flat potentials. This Multiverse can explain the values of the cosmological constant and of the weak scale. In the process of proving these statements we derive a few simple, but counter-intuitive results. We show that it is easy to write models of eternal inflation compatible with the distance and refined de Sitter conjectures. Secondly, tunneling transitions that move fields from a lower-energy vacuum to a higher-energy  vacuum and generate baby Universes are possible, and occur during eternal inflation. Finally, we relax the assumption of no de Sitter minima and show that this more standard Multiverse can be populated by Coleman-De Luccia transitions in about 100 $e$-folds of inflation.
\end{abstract}

\maketitle

%%%%%%%%%%%%%%%%%%%%%%%%%%%%%%%%%%%%%%%%%%%%%%%%%%%%%%%%%%%%%%%%%%%%%%%%%%%
%%%%%%%%%%%%%%%%%%%%%%%%%%%%%%%%%%%%%%%%%%%%%%%%%%%%%%%%%%%%%%%%%%%%%%%%%%%
\section{Introduction}
%%%%%%%%%%%%%%%%%%%%%%%%%%%%%%%%%%%%%%%%%%%%%%%%%%%%%%%%%%%%%%%%%%%%%%%%%%%
%%%%%%%%%%%%%%%%%%%%%%%%%%%%%%%%%%%%%%%%%%%%%%%%%%%%%%%%%%%%%%%%%%%%%%%%%%%
A vast program is underway to identify the Effective Field Theories (EFTs) that can be coupled to gravity. The bulk of these efforts focuses on finding general principles that allow to UV-complete an EFT without gravity into string theory. A series of ``Swampland conjectures"\footnote{See~\cite{Palti:2022edh} for a recent review.} was developed as part of this program. The conjectures are intended to identify all theories that live in the Swampland, as opposed to the landscape of healthy EFTs that can be derived from string theory. 

At the time of writing, the validity of these conjectures remains debated. The goal of this work is not to add to this debate, but rather to understand if any of these arguments can have an impact on open problems in fundamental physics. 

A particularly interesting question is whether a combination of the {\it distance}~\cite{Ooguri:2006in} and {\it refined dS}~\cite{Garg:2018reu, Ooguri:2018wrx} conjectures is in tension with slow-roll single-field inflation~\cite{Agrawal:2018own}, currently our best explanation for Cosmic Microwave Background (CMB) measurements. 

The tension with inflation suggests another possible tension between the two conjectures and a construction that is relevant to some of the biggest open problems in particle physics: the Multiverse. Before illustrating the technical results of the paper, it is useful to recall a few characteristics of the Multiverse:
  1) The physical processes that can create a Multiverse are simple and well known~\cite{Vilenkin:1983xq, Winitzki:2008zz, Guth:2007ng, Linde:1986fc, Linde:1986fd}, 2) The viability of this idea can be tested by searching for a UV completion of the SM with many metastable vacua and for a long period of inflation, 3) The Multiverse is the most concrete explanation that we have for the small value of the cosmological constant (CC), together with Weinberg's argument~\cite{Weinberg:1987dv}.

Other than for the CC, the Multiverse can play an important role also in a number of other open questions in fundamental physics, where the most notable example is the value of the Higgs mass squared ($m_h^2$)~\cite{Agrawal:1997gf, Arkani-Hamed:2004ymt, Geller:2018xvz, Giudice:2019iwl, Arkani-Hamed:2020yna, Strumia:2020bdy, Csaki:2020zqz, TitoDAgnolo:2021nhd, TitoDAgnolo:2021pjo, Giudice:2021viw,Khoury:2021zao}. 

Establishing if the Multiverse is consistent with what we know of string theory is of practical interest to particle physics. Theories that leverage the existence of a Multiverse to explain the Higgs mass have qualitatively different signatures compared to traditional symmetry-based explanations~\cite{Martin:1997ns, Panico:2015jxa}, and we would like to know in advance if building experiments that search for these signatures is worth it, or if the whole idea is inconsistent with quantum gravity.

It is of course possible that the Swampland conjectures do not accurately reflect the nature of quantum gravity. However, here we will show that the Multiverse can still exist even if they do. Additionally, it is legitimate to worry about the measure problem~\cite{Linde:1994gy,Vilenkin:1998kr,Guth:2007ng}, the existence of Boltzmann brains and other paradoxes encountered in eternal inflation when trying to explain measurements in our Universe. We do not add anything new to the solution of these problems, we just want to show that the Multiverse can exist outside of the Swampland. 

The Multiverse is traditionally associated with eternal inflation, which seems at odds with conjectured constraints on the EFT of inflation from string theory. In this work we derive a number of results that clarify the relation between the Multiverse and the Swampland: 
\begin{itemize}
\item Eternal inflation and a Multiverse can be realized in simple models of inflation with sub-Planckian field excursions and steep potentials ($M_{\rm Pl} |\nabla V|/V > 1$), without de Sitter (dS) minima. 
\item If the landscape is made up of Minkowski and Anti-de Sitter (AdS) minima, they can all be populated and thus create a Multiverse that explains the observed values of CC and $m_h^2$.
\item A Multiverse that scans both the CC and the weak scale does not require eternal inflation, it needs only $\mathcal{O}(100)$ $e$-folds if we allow dS minima.
\end{itemize}

%%%%%%%%%%%%%%%%%%%%%%%%%%%%%%%%%%
\begin{figure}[t]
\centering
\includegraphics[width=0.48\textwidth]{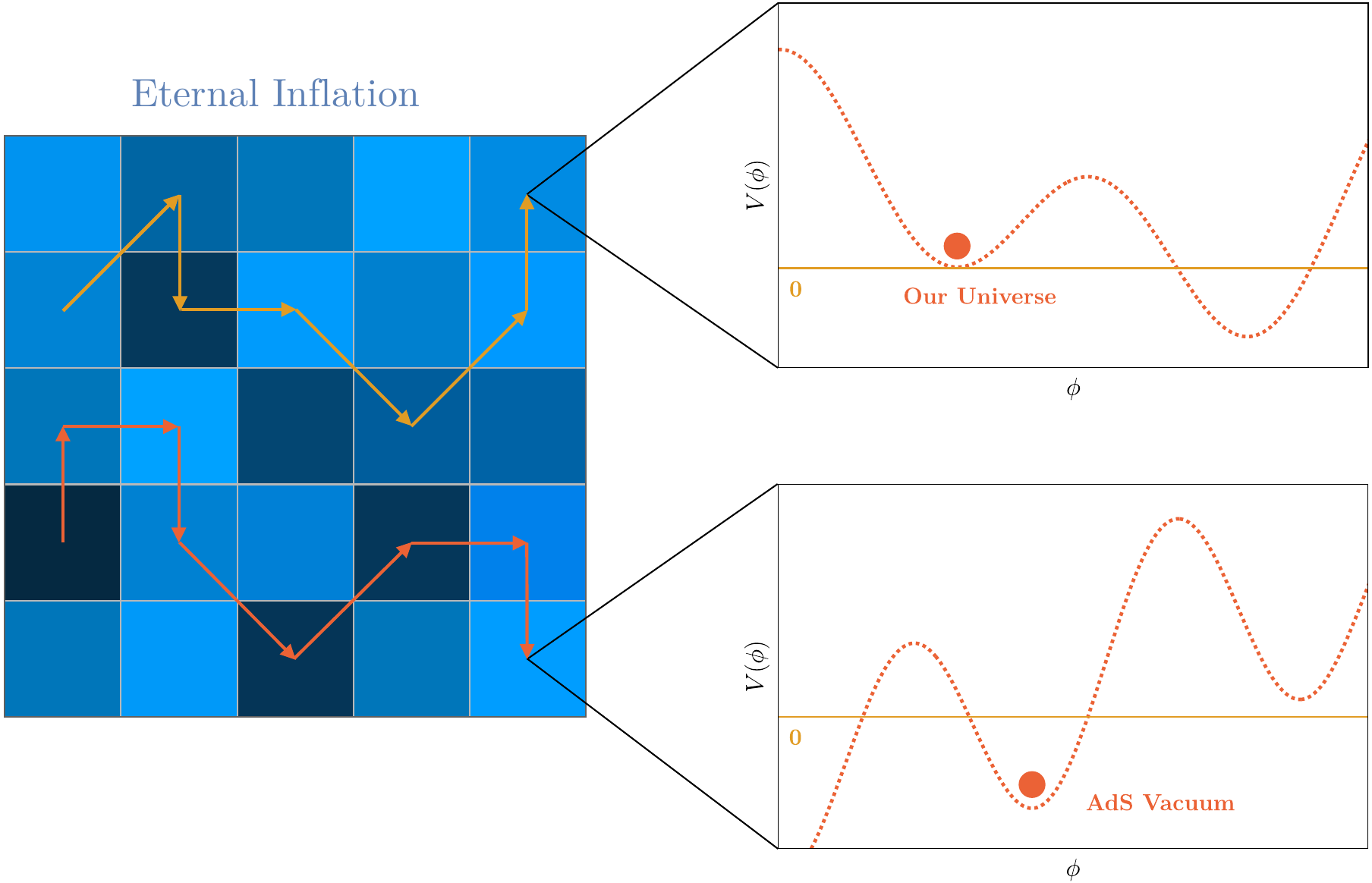}
\caption{A schematic drawing of the scenario under consideration. We begin with a period of eternal inflation-like expansion (blue region). Bubbles continue to be nucleated, with either higher or lower vacuum energies (represented by different shades). We assume there are only Minkowski and AdS minima in the landscape. As a region inside a bubble exits eternal inflation, it will either end up in an AdS minimum, or continue with 60 $e$-folds of slow-roll inflation and end up in a Minkowski minimum.}
\label{fig:scheme}
\end{figure}
%%%%%%%%%%%%%%%%%%%%%%%%%%%%%%%%%%

In addition to our main results on the Multiverse, we also briefly review the tension between slow-roll inflation (both the single-field and the multifield version) and the combination of distance conjecture and refined dS conjecture. We discuss how explicit models that are consistent with both conjectures can be constructed. In the process we also show that: 1) One cannot circumvent the conjectures by adding a single field to an existing inflationary model, 2) Warm inflation can explain the CMB in models with steep potentials, even though challenges remain to find a UV completion. 

Work related to all of our results was already published shortly after the conjectures were first proposed. We review the relation between our work and the existing literature in each of the physics Sections. 

The rest of the paper is organized as follows: in Section~\ref{sec:basic} we describe the basic setup that allows to populate a Multiverse compatible with the distance conjecture and refined dS conjecture, which we restate in this same Section.
In Section~\ref{sec:eternal} we write a simple model of eternal inflation compatible with both of  our working hypotheses. In Section~\ref{sec:Multiverse} we show that eternal inflation can populate a Multiverse with all allowed values for the CC and $m_h^2$ also if all minima in the landscape have only zero or negative CC. In Section~\ref{sec:CMB} we discuss how to explain CMB observations compatibly with our working hypotheses. In Section~\ref{sec:tunneling} we relax our second working hypothesis and show that the existence of a Multiverse with dS minima does not require more $e$-folds of inflation than explaining the CMB.

%%%%%%%%%%%%%%%%%%%%%%%%%%%%%%%%%%%%%%
\section{Basic Picture}\label{sec:basic}
To derive the results discussed in the introduction we consider the scenario shown in Fig.~\ref{fig:scheme}. We assume that there are only Minkowski and AdS minima in the landscape and that generic initial conditions place an observer in an AdS minimum with CC of $\mathcal{O}\left(-M_{\rm Pl}^4\right)$. We imagine that in a small patch of the Universe one or more fields are not in one of their minima. In this case, the vacuum energy is uplifted to be positive and starts eternal inflation. We discuss a simple model that supports eternal inflation near a maximum of a scalar potential in Section~\ref{sec:eternal}. 

If we start in a generic minimum of the landscape with CC of $\mathcal{O}\left(-M_{\rm Pl}^4\right)$, classically we might expect that we will still end up in the same place after inflation. However, during inflation, new bubbles with larger vacuum energies are nucleated and some of them expand into baby Universes. We show that tunneling transitions can generate Universes with a vacuum energy that is {\it larger} by $\mathcal{O}(1)$ in units of $M_{\rm Pl}^4$ than that of their parent Universe. Therefore, it is possible that eventually we generate an expanding patch that today has a tiny CC of $\mathcal{O}({\rm meV})^4$, compatible with the observed one. 

After exiting eternal inflation, the fates of the bubbles are different. Many of them end up in AdS minima, but at least one of them experiences 60 $e$-folds of slow-roll inflation and reaches a minimum very close to Minkowski,  which is our Universe.  We are rather agnostic on what explains today's observed dark energy: it could be quintessence~\cite{Peebles:1987ek,Ratra:1987rm,Frieman:1995pm,Ferreira:1997au,Caldwell:1997ii} or just a small violation of the refined dS conjecture (i.e.\ we are in a dS minimum with a tiny CC in units of $M_{\rm Pl}$). A recent observation by DESI favors a dark energy component that varies with time~\cite{DESI:2024mwx}, but the evidence is still far from conclusive.

In this way, a Multiverse is ``populated" and Weinberg's argument explains why we observe a small CC. If the fields in the landscape are also coupled to $|H|^2$ this same Multiverse can explain the value of $m_h^2$ via one of the ideas in~\cite{Agrawal:1997gf, Arkani-Hamed:2004ymt, Geller:2018xvz, Giudice:2019iwl, Arkani-Hamed:2020yna, Strumia:2020bdy, Csaki:2020zqz, TitoDAgnolo:2021nhd, TitoDAgnolo:2021pjo, Giudice:2021viw}. Of course, this is not the only possible way to realize the Multiverse and select our Universe. We pick this setup to illustrate our main physics points in the simplest possible way.
\subsection{Working Hypotheses}
Before getting to our main results we state the conjectures against trans-Planckian field excursions and shallow potentials that in this work we uplift to working hypotheses. From now on we only consider field excursions below $M_{\rm Pl}$,
\be
\Delta \phi \leq M_{\rm Pl}\, .
\label{eq:WP1}
\ee
Note that our definition of $M_{\rm Pl}$ is always $M_{\rm Pl}^2 \equiv (8\pi G_\tn{N})^{-1} = 2.39 \times 10^{18}$ GeV. 
The requirement on the field range that we adopt here is somewhat more stringent than the distance conjecture~\cite{Ooguri:2006in}, where an unknown $\mathcal{O}(1)$ parameter multiplies the right-hand side of the inequality. Additionally, we assume the so-called Refined de Sitter Conjecture~\cite{Ooguri:2018wrx, Garg:2018reu} to be true. The scalar potential of our theories coupled to gravity must satisfy at every point one of the two following conditions:
\be
|\nabla V| \geq \frac{c}{M_{\rm Pl}} V\, , \quad \text{or} \quad
\min\left(\nabla_i \nabla_j V\right)\leq - \frac{c^\prime}{M_{\rm Pl}^2} V\, ,
\label{eq:WP2}
\ee
with $c, c^\prime=\mathcal{O}(1)$. Note that explicit examples in string theory exist where $c \ll1$~\cite{Garg:2018reu}. For other arguments against dS minima in quantum gravity we refer to~\cite{Dvali:2013eja, Dvali:2014gua, Dvali:2017eba, Dvali:2020etd, Dvali:2021kxt}. 

%%%%%%%%%%%%%%%%%%%%%%%%%%%%%%%%%%%%%%%%%%%%%%%%%%%%%%%%%%%%%%%%%%%%%%%%%%%
%%%%%%%%%%%%%%%%%%%%%%%%%%%%%%%%%%%%%%%%%%%%%%%%%%%%%%%%%%%%%%%%%%%%%%%%%%%
\section{Eternal Inflation Outside of the Swampland}\label{sec:eternal}
%%%%%%%%%%%%%%%%%%%%%%%%%%%%%%%%%%%%%%%%%%%%%%%%%%%%%%%%%%%%%%%%%%%%%%%%%%%
%%%%%%%%%%%%%%%%%%%%%%%%%%%%%%%%%%%%%%%%%%%%%%%%%%%%%%%%%%%%%%%%%%%%%%%%%%%

Our two working hypotheses do not exclude eternal inflation. On the contrary, one can write a very simple model with large Hubble during inflation that can generate a Multiverse. Consider the hilltop inflationary potential
\be
V= m^2 f^2 \left(1-\frac{\phi^2}{2f^2}+\mathcal{O}(\phi/f)^4\right)\, , \label{eq:VEI}
\ee
that approximates a generic pseudo-Goldstone potential near a maximum\footnote{If there are other fields and they do not have dS minima, we should add to $V$ a negative CC. This complicates our discussion without changing it qualitatively, so we omit this constant in the following.}. Eternal inflation occurs if the quantum fluctuations of the field dominate over its classical motion. In this limit, some spacetime patches never leave the region of the potential that drives inflation, giving rise to an ever-expanding Multiverse~\cite{Vilenkin:1983xq, Winitzki:2008zz, Guth:2007ng, Linde:1986fc, Linde:1986fd}. 

In one Hubble time the classical motion of the field is
\be
\Delta \phi_\tn{c} = \frac{V^\prime}{a H^2}\, , 
\ee
where $a$ is an $\mathcal{O}(1)$ number. In the case of slow-roll, $a=3$. Over the same timescale, the field experiences quantum fluctuations with standard deviation
\be
\Delta \phi_\tn{q} = \frac{H}{2\pi}\, .
\ee
To have eternal inflation we need $\Delta \phi_\tn{q} \gtrsim \Delta \phi_\tn{c}$ or equivalently
\be
V^\prime \lesssim \frac{a H^3}{2\pi}\, . \label{eq:EI}
\ee
Given how vague the conjectures are and the approximations involved in this inequality, it seems foolish to keep the $\mathcal{O}(1)$ number $a$ or even the factor of $2\pi$. However, in the discussion around Eq.~\eqref{eq:prob} it will be clear why we did it. 

%%%%%%%%%%%%%%%%%%%%%%%%%%%
\begin{figure}[!t]
\centering
\includegraphics[width=0.48\textwidth]{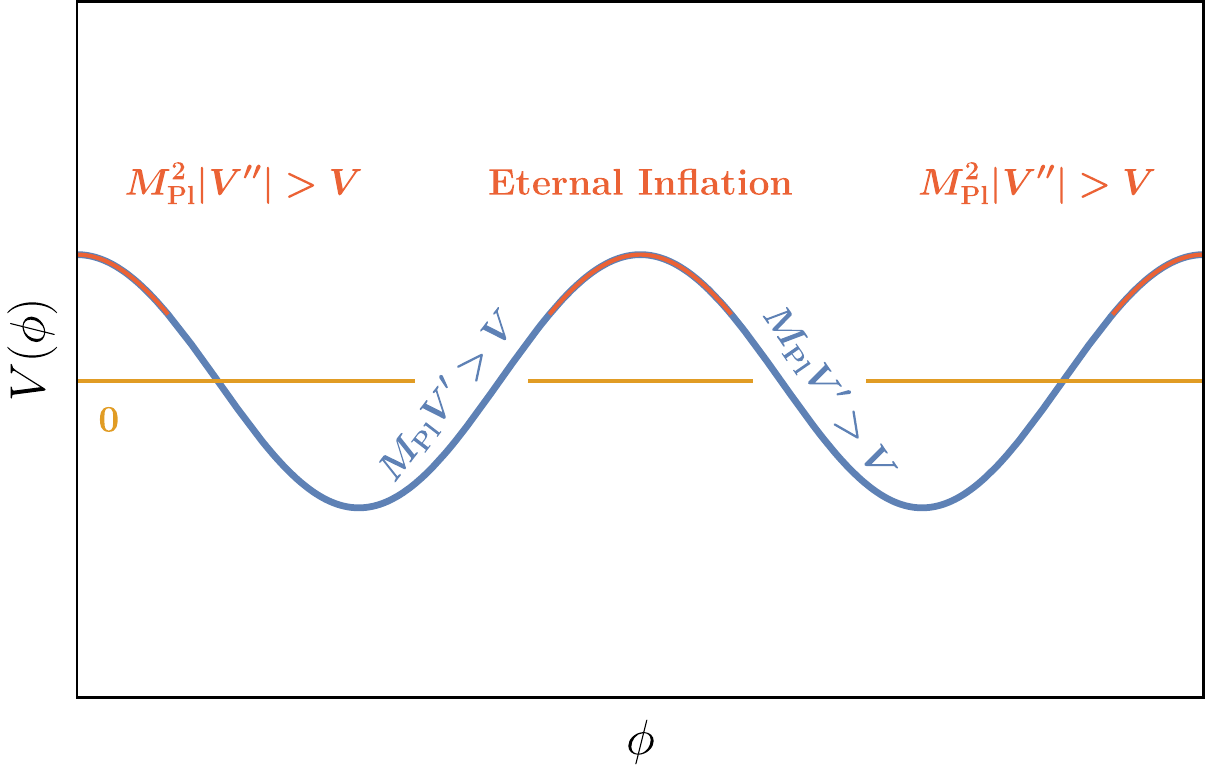}
\caption{A model of eternal inflation compatible with Swampland conjectures. A pseudo-Goldstone boson, with decay constant and mass just below $M_{\rm Pl}$, is close to its maximum. Near the top of the potential, the second inequality in the de Sitter conjecture is satisfied, while the first inequality is satisfied elsewhere. The minima have zero or negative CC. Eternal inflation requires a mild tuning of initial conditions described in the text.}
\label{fig:EI}
\end{figure}
%%%%%%%%%%%%%%%%%%%%%%%%%%%%

From Eq.~\eqref{eq:EI} we see that it is possible to realize eternal inflation if we accept a mild tuning of initial conditions for the position and velocity of the field:
\be
\frac{\phi_\tn{I}}{f}\lesssim \delta\, , \quad \frac{\dot \phi_\tn{I}}{m f} \lesssim \delta \frac{H}{m} \simeq \delta \frac{f}{M_{\rm Pl}} \, ,  \label{eq:eternal}
\ee
where
\be
\delta \equiv \frac{a}{6 \sqrt{3}\pi}\frac{m f^2}{M_{\rm Pl}^3} \lesssim 1\, .
\ee
The initial position $\phi_\tn{I} \lesssim \delta f$ puts us close enough to the top of the potential for quantum fluctuations to dominate over classical rolling. The velocity in Eq.~\eqref{eq:eternal} is small enough to move $\phi$ less than $\delta f$ during one Hubble time ($\dot \phi \lesssim \delta f H$).
The total tuning is then of order
\be
\frac{\phi_\tn{I}}{f} \frac{\dot \phi_\tn{I}}{m f}\simeq \delta^2 \frac{f}{M_{\rm Pl}}\, ,
\ee
since the natural value for the position of the field is $f$ and for its velocity is $m f$. This tuning can be extremely mild because nothing stops us from taking $m$ and $f$ very close to  $M_{\rm Pl}$. 

\begin{figure*}[!t]
\includegraphics[width=0.49\textwidth]{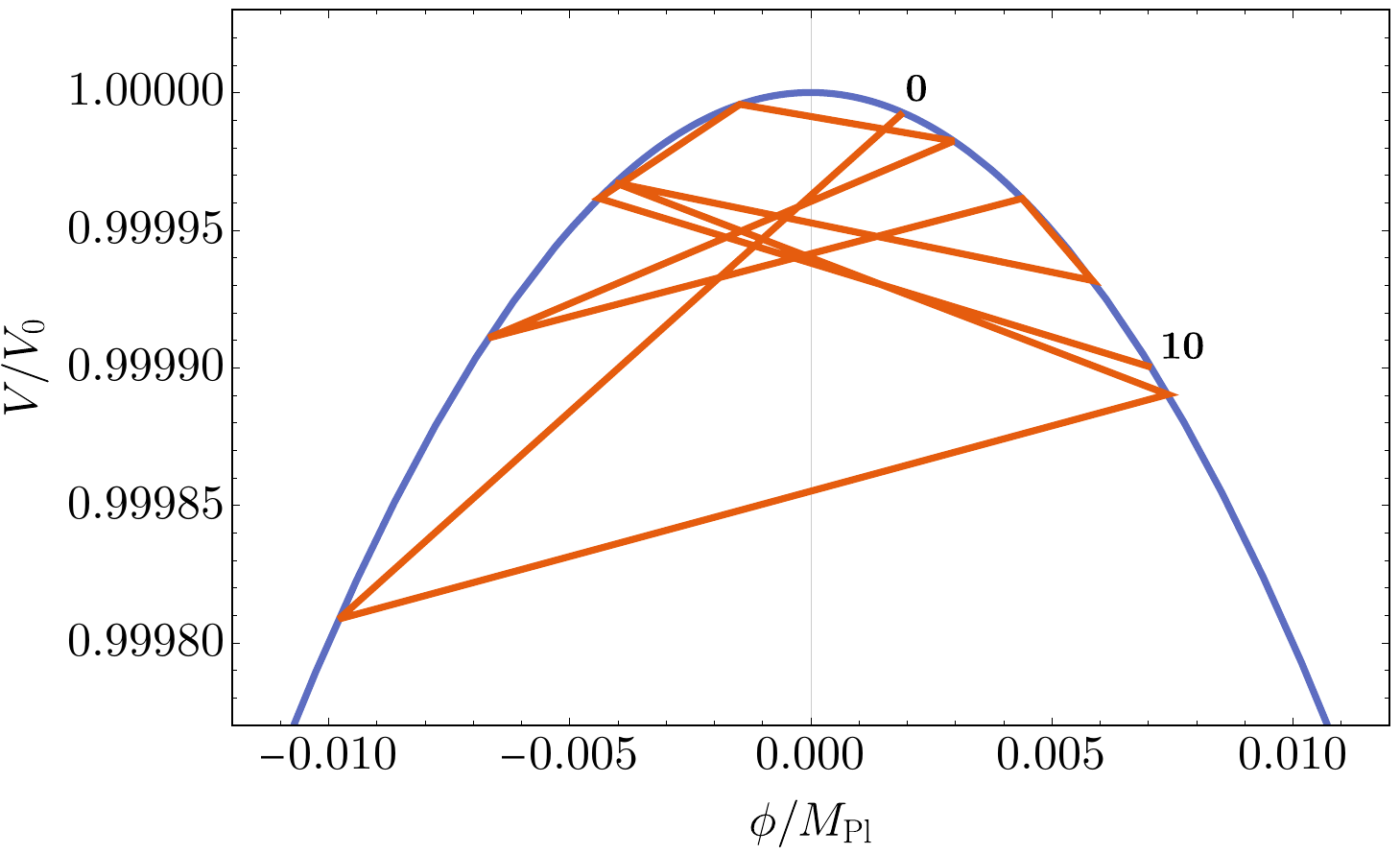}\quad
\includegraphics[width=0.49\textwidth]{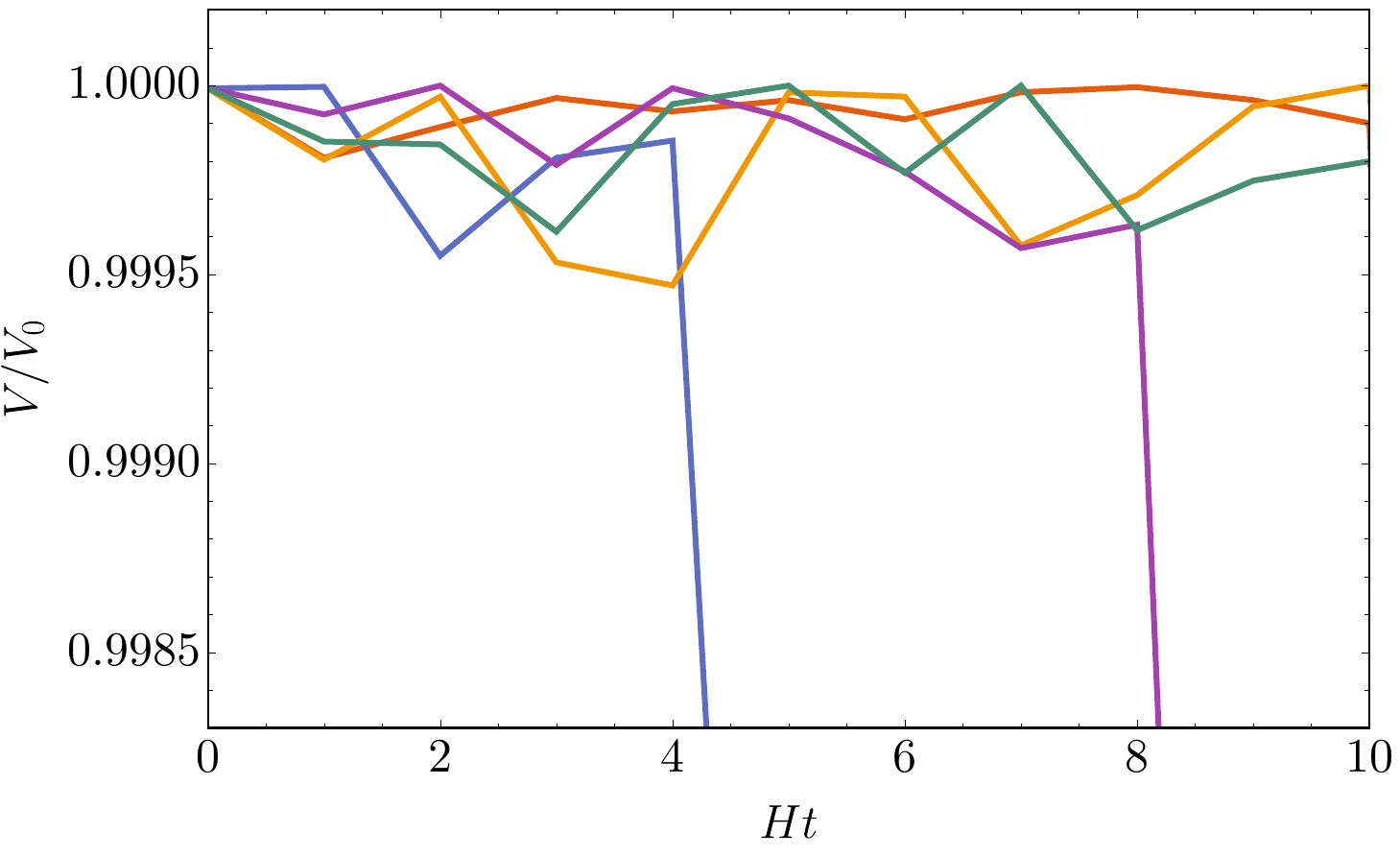}
\caption{Dynamics of an inflaton in the hilltop potential of Eq.~\eqref{eq:VEI}. Time is discretized in units of $H^{-1}$ and at each time step quantum fluctuations are generated by sampling a Gaussian distribution. The regime in which quantum fluctuations dominate over classical motion can be achieved by properly tuning the initial conditions for position and velocity. Then in some Hubble patches the field remains close to the top of the potential forever. On the left we show an example of such a case, in which the inflaton is very close to the top after 10 Hubble times (we stop the plot here for simplicity, since we are already sampling $e^{30}$ patches). On the right we show several trajectories, including two that fall and exit the eternal inflation regime after a few Hubble times. The plots are made by starting with a single Hubble patch near the top of the potential and following the evolution of the $e^{3H t}$ patches generated at every step.}
\label{fig:eternal}
\end{figure*}

The refined dS conjecture in Eq.~\eqref{eq:WP2} is respected for $\phi$ away from the maximum if $c f\leq M_{\rm Pl}$. At the maximum, where $V^\prime=0$, we have to satisfy the second inequality in Eq.~\eqref{eq:WP2},
\be
-\frac{M_{\rm Pl}^2V^{\prime\prime}}{V}=\frac{M_{\rm Pl}^2}{f^2}\geq c^\prime\, . 
\ee
If $f\leq M_{\rm Pl}$ (as implied by the above inequalities for $c, c^\prime = \mathcal{O}(1)$) our stronger version of the distance conjecture in Eq.~\eqref{eq:WP1} is automatically satisfied. We assume that the UV completion of Eq.~\eqref{eq:VEI} does not have dS minima, a simple possibility is $V=m^2 f^2 \cos(\phi/f)$. This example potential which satisfies all conjectures is shown in Fig.~\ref{fig:EI}.

Note that the ``large" second derivative of the potential does not destroy eternal inflation, as the velocity of the field changes only at $\mathcal{O}(1)$ during each Hubble time,
\be
\frac{\Delta \dot \phi}{\dot \phi} \simeq - \frac{3 c^\prime}{a}\, ,
\ee
and this variation can change sign, as quantum fluctuations will in general move the field from one side of the maximum to the other.

The last point to consider is the size of the region in field space that supports eternal inflation. If $\phi_\tn{I}$ in Eq.~\eqref{eq:eternal} is small,  
\be
x_\tn{I} \equiv \frac{2\pi \phi_\tn{I}}{H} \simeq \frac{a}{3}\frac{f^2}{M_{\rm Pl}^2} < 1 \label{eq:xI}\, ,
\ee
we need an additional accident  to realize eternal inflation, because we do not want to exit this region in the first few steps of the quantum random walk of the field. While presenting this argument we assume, conservatively, that inflation starts in a single Hubble-sized patch. This is usually not the case, but since we are tuning to be close to the top of the potential, which makes starting inflation more unlikely, we want to consider also this potential worst-case scenario.

The probability that one step leaves us in the region $|\phi|\lesssim \phi_\tn{I}$ is
\be
p=\int_{-x_\tn{I}}^{x_\tn{I}} dx \frac{e^{-\frac{x^2}{2}}}{\sqrt{2\pi}}\, . \label{eq:prob}
\ee
This equation is why we kept the $\mathcal{O}(1)$ number $a$ and the factor of $2\pi$. Eq.~\eqref{eq:prob} shows that they are exponentiated in the measurement of this potential extra tuning. 

For concreteness we can first consider specific values of $m, f$ that give $p\gtrsim 1/20$. We can model the quantum motion of the field as sampling, every Hubble time, a Gaussian distribution of zero mean and standard deviation $H/2\pi$. If initially $|\phi|\lesssim \phi_\tn{I}$ in a region of size $H^{-1}$ and our accident occurs once, then at the second step of the random walk the Universe has already expanded and its volume is $e^3$ times bigger. Since dS space has a horizon of size $H^{-1}$, we need to sample the probability distribution separately in twenty regions of size $H^{-1}$, making the overall probability that one of these regions keeps inflating $\mathcal{O}(1)$. 
Following this logic it is easy to show that for a generic value of $p$ the overall tuning is\footnote{We assumed a discrete time evolution in steps of $\Delta t \simeq H^{-1}$ and that the scalar starts moving before the first step.}  
\be
\sim p \prod_{i=1}^{N_e} \sum_{n_i=1}^{n_{i-1} e^3} B\left(n_i, n_{i-1} e^3\right)\, , \quad n_0=1\, ,
\ee
 where $B(n, N)\equiv \left(\begin{array}{c}N \\ n\end{array}\right) p^n (1-p)^{N-n} $ is the usual binomial distribution and $N_e$ is the number of $e$-folds (steps in the random walk). Once again, the probability of this accident can be large, since we can take $f$ in Eq.~\eqref{eq:xI} close to $M_{\rm Pl}$. In Fig.~\ref{fig:eternal}, we show some examples of the evolution of an inflaton for about 10 $e$-folds. 

%%%%%%%%%%%%%%%%%%%%%%%%%%%
\begin{figure*}[!t]
\centering
\includegraphics[width=\textwidth]{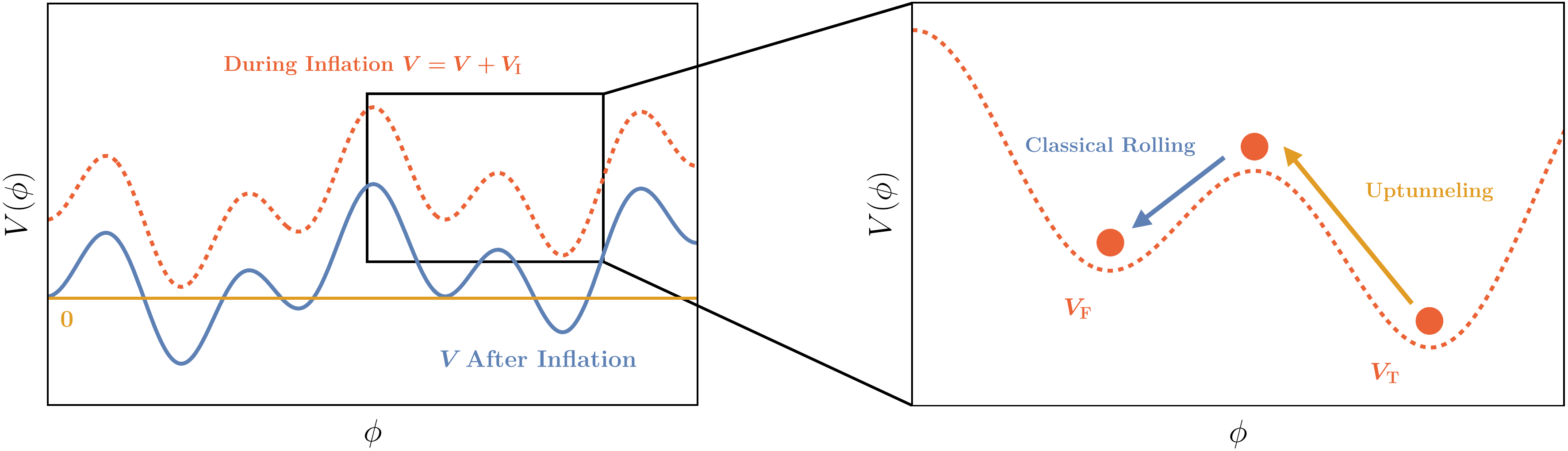}
\caption{Schematic view of the processes that populate a Minkowski+AdS Multiverse. During inflation an $\mathcal{O}(1)$ fraction of minima in the landscape is in de Sitter space (left panel). In de Sitter space, a number of processes allow to tunnel to vacua of higher energy. In the right panel of the Figure we depict the effect of a Hawking-Moss instanton. In the text we show that these bubbles of false vacuum can generate baby Universes.}
\label{fig:up}
\end{figure*}
%%%%%%%%%%%%%%%%%%%%%%%%%%%%

We now have a simple model of eternal inflation compatible with our two working hypotheses in Eqs.~\eqref{eq:WP1} and~\eqref{eq:WP2}. However, this model alone does not describe our Universe, as was noted for instance in~\cite{Wang:2019eym}. What we imagine is that a second inflaton takes care of the last 60 $e$-folds of inflation, as discussed in Section~\ref{sec:CMB}. Here, we comment on why the simple model discussed in this Section is at odds with CMB measurements. 

If we imagine that the last 60 $e$-folds occur while $\phi$'s motion is dominated by quantum fluctuations, the predicted power spectrum is flat~\cite{Kleban:2012ph},
\be
\Delta_R^2 \simeq \frac{1}{(2\pi)^2}\, ,
\ee
and the CMB quadrupole is too large~\cite{Zeldovic_Grisuchuk}. Unfortunately we cannot imagine to exit inflation via slow-roll, because slow-roll requires to be exponentially close to the top of the potential where quantum fluctuations dominate. On this potential, slow-roll occurs for $N_e$ $e$-folds only if initially\footnote{In a more generic potential $V=\Lambda^4-m^2 \phi^2/2$, the tuning is usually worse. If all other fields are at their (AdS or Minkowski) minimum then $\Lambda^4 < m^2 f^2$ and $\phi_\tn{I}(N_e)=\frac{\Lambda^4}{m^2 M_{\rm Pl}} e^{-\frac{m^2 M_{\rm Pl}^2}{\Lambda^4} N_e}$.}
\be
\frac{\phi_\tn{I}(N_e)}{f}=\frac{f}{M_{\rm Pl}} e^{-\frac{M_{\rm Pl}^2}{f^2} N_e}\, . \label{eq:slowtuning}
\ee
Hence, the hilltop potential alone does not describe our Universe. However,  this is not a problem if we are willing to imagine a second inflationary sector which is responsible for the last 60 $e$-folds and explains the CMB, after our Universe exits eternal inflation. We discuss this possibility in Section~\ref{sec:CMB}. 

To conclude this Section we comment on related work. First of all, eternal inflation can be excluded by a future measurement of positive curvature~\cite{Kleban:2012ph} that would exclude an epoch of eternal inflation followed by slow-roll, or by a negative running of the spectral index~\cite{Kinney:2014jya,Montefalcone:2023izs}. Planck is not sensitive enough to make a definitive statement~\cite{Planck:2018jri}.

Related analyses in the literature can be found in~\cite{Wang:2019eym, Dimopoulos:2018upl, Matsui:2018bsy, Blanco-Pillado:2019tdf,Kinney:2018kew,Brahma:2019iyy,Lin:2019fdk}. In~\cite{Wang:2019eym} the same model is considered and discarded because of the difficulties that we mentioned in explaining the CMB. In~\cite{Dimopoulos:2018upl} they consider a linear potential near the top and do not tune, concluding that eternal hilltop inflation is incompatible with the refined dS conjecture. In~\cite{Matsui:2018bsy} they consider eternal chaotic inflation and derive bounds on the $\mathcal{O}(1)$ parameters of the conjectures to accommodate it. It is possible only if, in Eq.~\eqref{eq:WP2}, $c\lesssim 0.01$ and the field excursion is $\gtrsim 10^2M_{\rm Pl}$.  Finally, in~\cite{Blanco-Pillado:2019tdf} an alternative way to have eternal inflation in a landscape without dS minima is discussed.

%%%%%%%%%%%%%%%%%%%%%%%%%%%
\begin{figure*}[!t]
\includegraphics[width=0.49\textwidth]{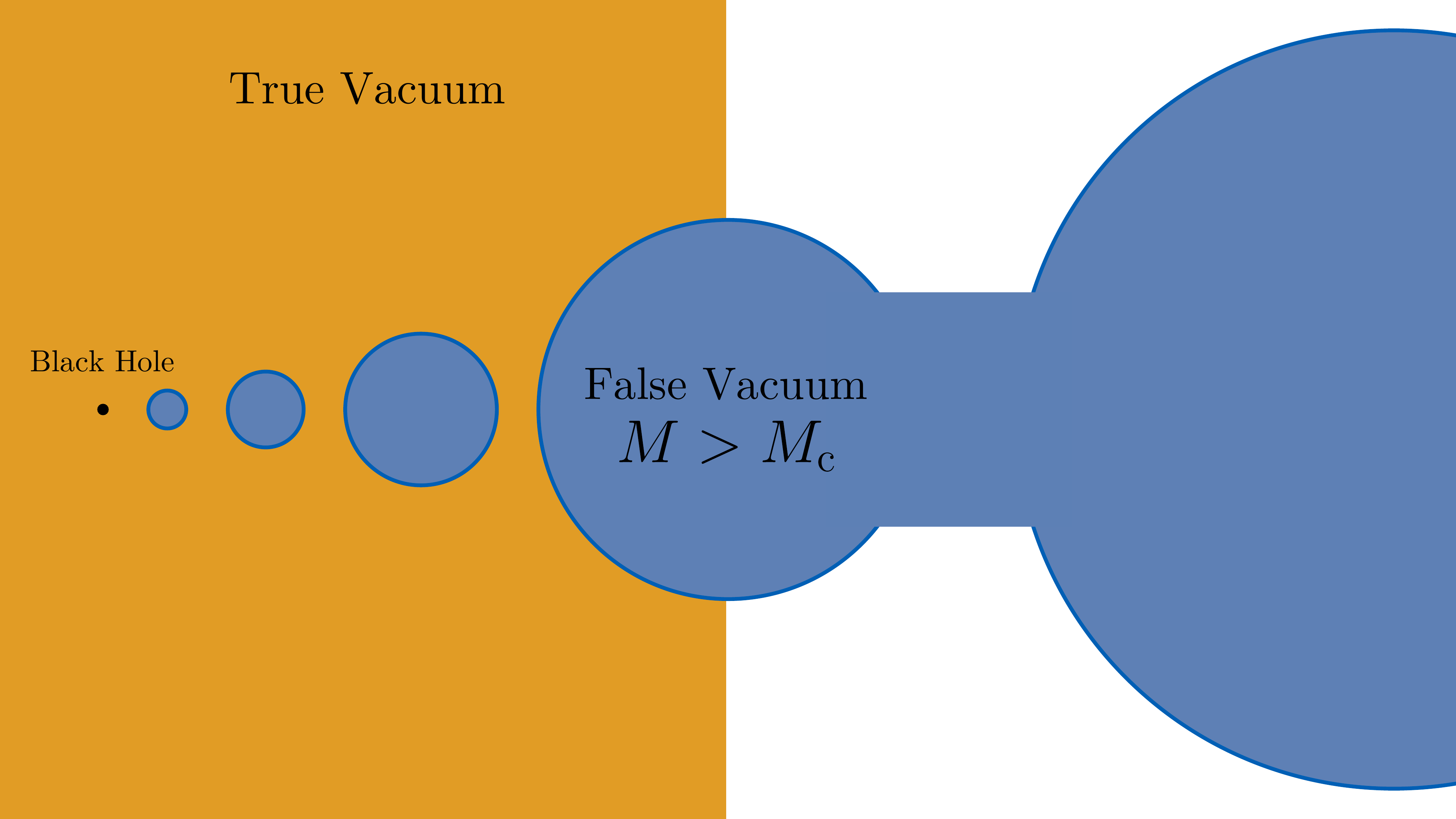}\quad
\includegraphics[width=0.49\textwidth]{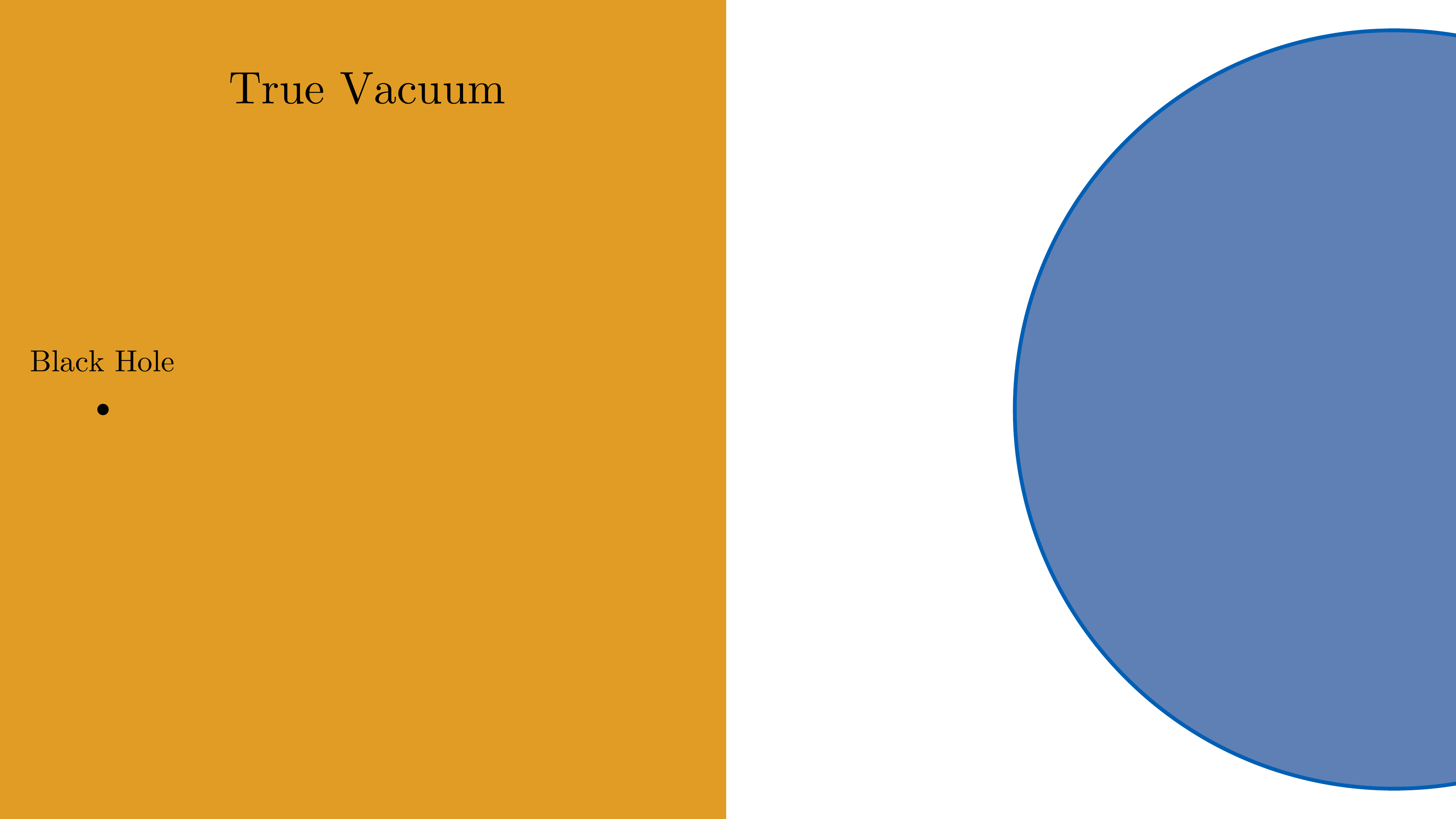}
\caption{Dynamics of large ($M>M_\tn{c}$) false vacuum bubbles in dS. The observer in the true vacuum sees the bubble shrink to a black hole (left panel). The observer inside the bubble sees an exponential expansion of spacetime that generates a baby Universe. After a while the throat connecting the bubble to the parent Universe pinches off (right panel) and a baby Universe in the false vacuum is born.}
\label{fig:bubble}
\end{figure*}
%%%%%%%%%%%%%%%%%%%%%%%%%%%%

Another potential issue with the eternal inflation scenario is the so-called ``trans-Planckian problem"~\cite{Martin:2000xs}, which was recently promoted to a Trans-Planckian Censorship Conjecture~\cite{Bedroya:2019tba,Bedroya:2019snp}. Requiring modes with sub-Planckian size to not get stretched beyond the horizon during inflation puts a limit on the number of $e$-folds, which would further constrain the eternal inflation scenario under consideration here. However, this requirement is not a necessary condition to maintain the validity of the effective field theory which describes inflation~\cite{Burgess:2020nec,Komissarov:2022gax}. We will not consider this constraint further here. 

%%%%%%%%%%%%%%%%%%%%%%%%%%%%%%%%%%%%%
\section{A Multiverse Without de Sitter Minima}\label{sec:Multiverse}
We have seen that eternal inflation is consistent with both our working hypotheses Eqs.~\eqref{eq:WP1} and~\eqref{eq:WP2}. But this is not enough to have a Multiverse that explains the CC and the weak scale, since we consider all minima in the landscape to be Minkowski or AdS, and their typical energy density to be $-M_{\rm Pl}^4$. In this landscape, natural initial conditions place us in a minimum with a CC of $\mathcal{O}\left(-M_{\rm Pl}^4\right)$. During the eternal inflationary period, we need to get to a minimum with higher vacuum energy in order to describe our Universe. Similarly, a typical minimum has $m^2_h \simeq M_{\rm Pl}^2$ and we need many (up)tunneling events to get to our Universe.

In this Section we show how to scan the CC in this landscape. If $|H|^2$ is coupled to the fields that generate the landscape also $m_h^2$ scans at the same time. We imagine to be in the eternal inflation regime described in the previous Section. An inflaton $\phi$ dominates the energy density of the Universe, so long as it stays close to the top of its potential. Other fields that we do not specify populate a vast landscape of vacua. For simplicity we take all of these fields in one of their local minima. When $\phi$ relaxes to its own minimum, all the local minima in the landscape are in AdS or Minkowski, but during inflation an $\mathcal{O}(1)$ fraction of them lives in dS space.

In dS one can always tunnel beyond the top of the potential separating two minima, via the Hawking-Moss (HM) instanton~\cite{Hawking:1981fz} or via a ``flyover" transition, where a field acquires a large kinetic energy from quantum fluctuations during inflation~\cite{Blanco-Pillado:2019xny, Blanco-Pillado:2019tdf}. The field then rolls to the minimum on the other side of the potential barrier. This setup is represented schematically in Fig.~\ref{fig:up}.

In Section~\ref{sec:wall} we describe the general dynamics of a bubble of false vacuum inside a Universe in the true vacuum and derive the necessary condition for the bubble to expand and form another Universe. We base our results on the work in~\cite{Lee:1987qc, PhysRevD.36.2919, BEREZIN198323, Berezin:112638, Basu:1991ig, Sato:1981gv, Blau:1986cw}. Initially we do not specify the nucleation mechanism. In Section~\ref{sec:HM} we consider bubbles nucleated by the HM instanton and show that they can create new Universes.

%%%%%%%%%%%%%%%%%%%%%%%%%%%%%%%%%
\subsection{Bubble Wall Dynamics}\label{sec:wall}

%%%%%%%%%%%%%%%%%%%%%%%%%%%%%%%%%
Inside the bubble we have a dS space in the false vacuum:
\beq
ds_{\rm inside}^2=-(1-H_\tn{F}^2 r^2) dt^2 + \frac{dr^2}{(1-H_\tn{F}^2 r^2)}+r^2 d\Omega_2^2\,.
\eeq
Outside the bubble we can use Birkhoff's theorem to write a dS-Schwarzchild metric~\cite{Ellis:2013dla}
\beq
\begin{aligned}
ds^2_{\rm outside} &=- f(\rho) dt^2 + \frac{d\rho^2}{f(\rho)}+\rho^2 d\Omega_2^2\, , \\
f(\rho)&= 1 - \frac{2 G_\tn{N} M}{\rho}- H_\tn{T}^2 \rho^2\, .
\end{aligned}
\eeq
Here $M$ is the total energy of the bubble that can be derived from its equation of motion, as we show below, and $H_\tn{F,\,T}=\sqrt{8\pi G_\tn{N} V_\tn{F,\,T}/3}$ are the Hubble parameters of the false and true vacua, $H_\tn{F} \geq H_\tn{T}$. In this Section we consider $M, H_\tn{F,\,T}$ and the energy per unit area of the bubble wall $\sigma$, introduced in the following, as free parameters.

Following~\cite{Blau:1986cw} one can derive an equation of motion for the radius of the wall $r$ from Israel's junction conditions~\cite{Israel:1966rt}
\be
\beta_\tn{F}-\beta_\tn{T}=4\pi G_\tn{N} \sigma r\, , \label{eq:master}
\ee
where we have defined
\beq
\begin{aligned}
\beta_\tn{F}&\equiv\pm\sqrt{1-H_\tn{F}^2 r^2+\dot r^2}\, ,\\
\beta_\tn{T}&\equiv\pm\sqrt{1-H_\tn{T}^2r^2-\frac{2G_\tn{N} M}{r}+\dot r^2}\, ,
\end{aligned}
\eeq
and $\sigma$ is the bubble energy per unit area. For a scalar $\phi_\tn{t}$ that is tunneling it reads~\cite{Coleman:1977py, Coleman:1980aw} 
\be
\sigma = \int_{\phi_\tn{I}}^{\phi_\tn{F}} d \phi_\tn{t} \sqrt{2 V(\phi_\tn{t})}\, .
\ee

%%%%%%%%%%%%%%%%%%%%%%%%%%%
 \begin{figure*}[!t]
\includegraphics[width=0.49\textwidth]{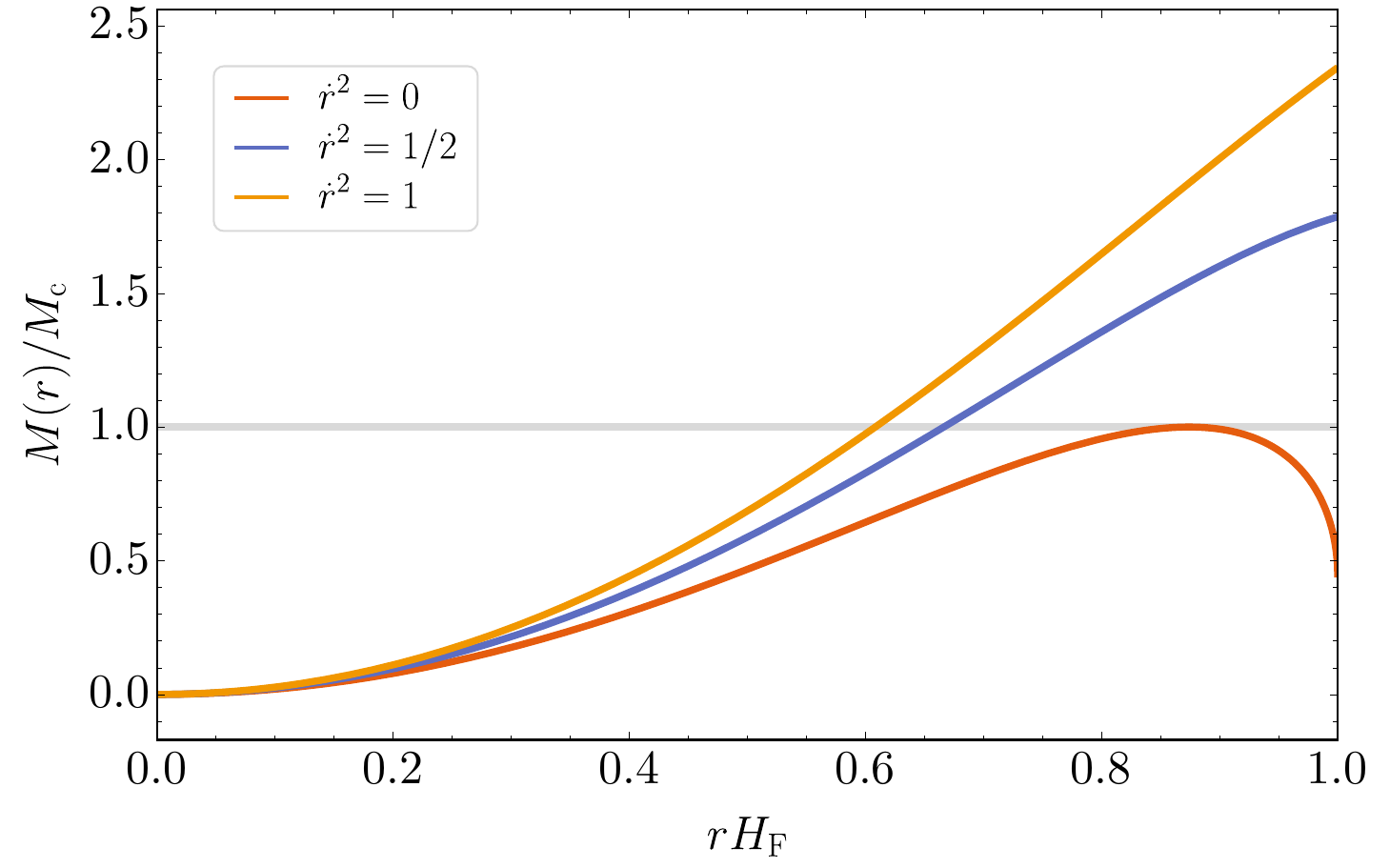}\quad
\includegraphics[width=0.49\textwidth]{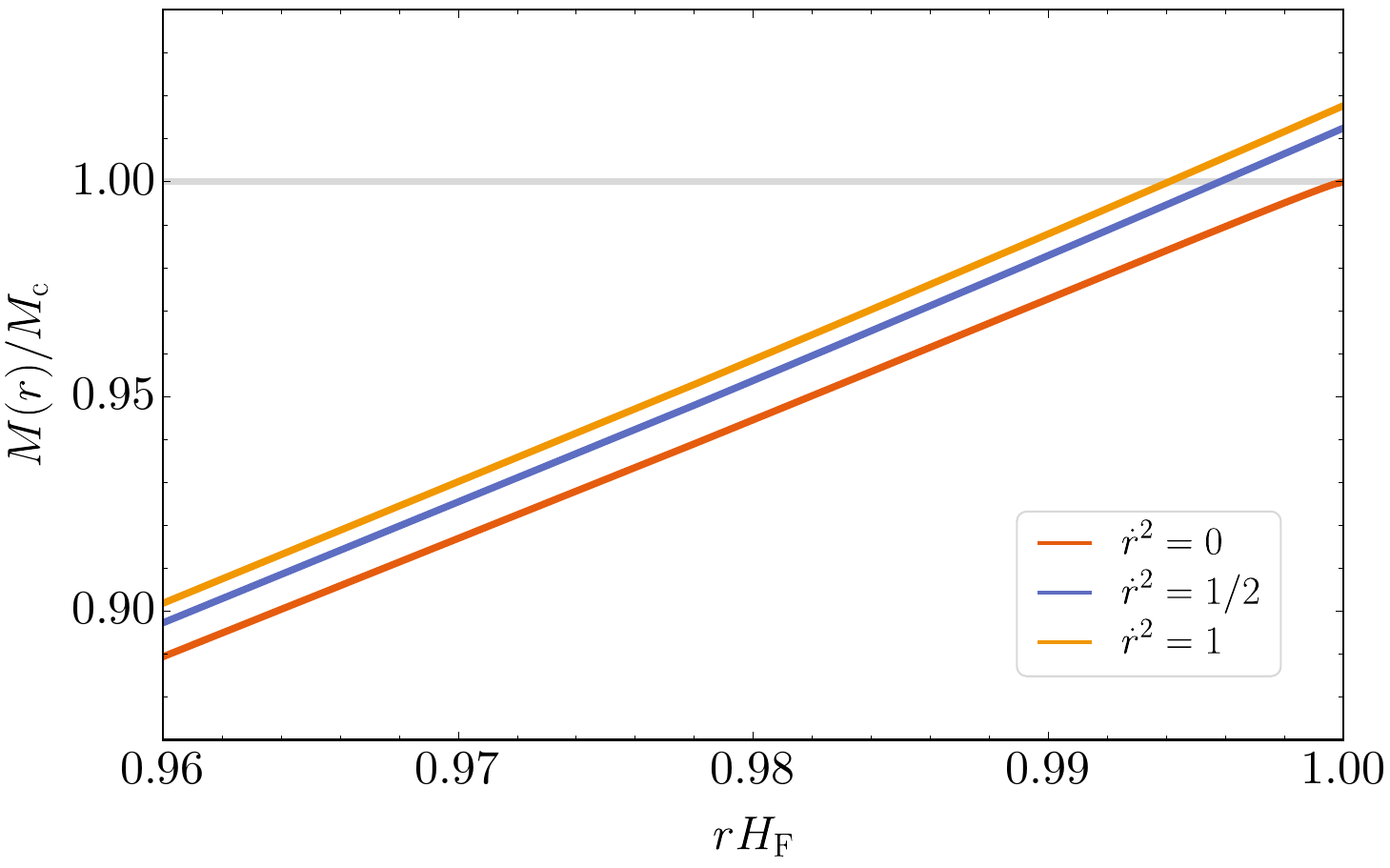}
\caption{Ratio of $M(r)$ to the critical bubble mass $M_\tn{c}$ for different values of $\dot{r}$. On the left, we pick the representative values $H_\tn{F}=2H_\tn{T}=3H_\sigma/2=M_\tn{Pl}$. On the right, the typical scale of our potential is $M_*=M_{\rm GUT}$, and so $H_\sigma\ll H_\tn{F,\,T}$. In both cases, we find that every $\dot{r}\neq0$ can give rise to supercritical bubbles consistent with causality.}
\label{fig:ratio}
\end{figure*}
%%%%%%%%%%%%%%%%%%%%%%%%%%%%

In the above equations a dot is a derivative with respect to $\tau$, the proper time of an observer moving with the wall. From Eq.~\eqref{eq:master} we can derive the mass of the bubble 
\beq
\begin{aligned}
M&=\frac{r^3 (H_\tn{F}^2-H_\tn{T}^2)}{2 G_\tn{N}}+ 4\pi \sigma r^2 \sqrt{1-H_\tn{F}^2 r^2+\dot r^2}{\rm sign}(\beta_\tn{F}) \\
&\phantom{{}={}}-8\pi^2 G_\tn{N} \sigma^2 r^3\, . \label{eq:mass}
\end{aligned}
\eeq
Following the same procedure detailed in~\cite{Blau:1986cw} it is easy to show that Eq.~\eqref{eq:master} describes the birth of a new Universe for bubbles with a mass above a certain critical value $M_\tn{c}$. If we define
\be
H_\sigma\equiv4\pi G_\tn{N}\sigma\, , \quad \Delta H^2_\pm \equiv H^2_\tn{F}-H^2_\tn{T}\pm H^2_\sigma\, , \label{eq:def}
\ee
the critical mass can be written as
\be
M_\tn{c}=\frac{R^3_\tn{c}\Delta H^2_-}{2G_\tn{N}}+\frac{R_\tn{c}H^2_\sigma \left(6R^2_\tn{c}H^2_\tn{F}-4\right)}{3G_\tn{N}\Delta H^2_-}\, , \label{eq:Mcrit}
\ee
where $R_\tn{c} \equiv \arg \max_r M(r, \dot r=0, \beta_\tn{F}>0)$. The derivation of $M_\tn{c}$ can be found in Appendix~\ref{app:mass}.  

For $M>M_\tn{c}$ we have a solution of Einstein's equations where an observer inside the bubble sees $r$ grow without bounds, while an observer outside sees the bubble collapse to a black hole. A careful spacetime analysis\footnote{Note that the analysis in~\cite{Blau:1986cw} applies to us when $M_\tn{c} > M_\tn{T} > M_\tn{F}$ where $M_\tn{F,\,T}$ are the masses for which $\beta_\tn{F,\,T}$ change sign. For us this condition means $H_\tn{F}^2-H_\tn{T}^2 > H_\sigma^2$.}~\cite{Blau:1986cw} shows that a baby Universe is created by this process. Initially, the inside of the bubble expands exponentially. Soon after the bubble formation and initial expansion, the throat connecting the inside to its parent Universe pinches off, leaving a black hole on the side of the true vacuum and a disconnected baby Universe that continues to expand on the side of the false vacuum. This is shown schematically in Fig.~\ref{fig:bubble}.

In Fig.~\ref{fig:ratio} we show $M(r)/M_\tn{c}$ as a function of $r$ for different values of $\dot r$. We find that supercritical bubbles can be nucleated only if $\dot r^2 >0$.

%%%%%%%%%%%%%%%%%%%%%%%%%%%%%%%%%
\subsection{Dynamics of a Hawking-Moss Bubble}\label{sec:HM}
First of all, we note that producing a baby Universe requires a quantum tunneling process~\cite{Farhi:1989yr, Fischler:1989se}, and classical production leads to a singularity in the past lightcone of the new Universe~\cite{Farhi:1986ty}. The theorem in~\cite{Farhi:1986ty} technically does not apply to Universes generated in dS space, since dS space has compact Cauchy surfaces, but the only loophole that this leaves open is producing the baby Universe everywhere in the parent Universe at once\footnote{We thank R. Wald for clarifying this point.}.

Therefore we consider quantum tunneling processes as the origin of our Multiverse. There are several ways to tunnel from a lower energy vacuum at $V_\tn{T}$ to a higher energy one at $V_\tn{F}$, if we are initially in a space with positive CC. The best-known one is through a Hawking-Moss instanton~\cite{Hawking:1981fz} that creates a vacuum bubble sitting at the top of the potential barrier separating the two minima. The field can then roll classically to its higher energy minimum $V_\tn{F}$. The HM instanton is an $O(4)$ symmetric solution of Einstein's equations with metric
\beq
\begin{aligned}
ds^2&= d\eta^2+\rho(\eta)^2 d\Omega_3^2\, , \\
\rho(\eta)&=H_{\rm top}^{-1} \sin (\eta H_{\rm top}) \, , 
\end{aligned}
\eeq
where we defined the Hubble parameter at the top of the potential in analogy with $H_\tn{F,\,T}$. The typical size of a HM bubble is simply~\cite{Hawking:1981fz}
\be
R_{\text{HM}}=\frac{1}{H_{\rm top}}\, .
\ee
The tunneling rate can be written in the same form as the Coleman-De Luccia one~\cite{Hawking:1981fz}:
\beq
\begin{aligned}
\Gamma &\simeq V M_*^4 e^{-S_\tn{HM}}\, , \\ 
S_{\rm HM} &= 8 \pi^2 M_{\rm Pl}^4 \left(\frac{1}{V_\tn{T}}-\frac{1}{V_{\rm top}}\right)\, .
\end{aligned}
\eeq
This instanton can be visualized as a series of quantum fluctuations of the field value occurring in dS space. The field always goes in the same direction in steps of $\sim H/2\pi$ until it finds itself at the top of the potential and then rolls to the other side. This also intuitively explains the exponentially small probability of this configuration. The quantum fluctuations in dS are approximately Gaussian with standard deviation $H/2\pi$ and the probability that multiple fluctuations occur in the same direction is proportional to  $\sim e^{-\frac{4 \pi^2 M_{\rm Pl}^2}{H^2}}$. 

\begin{figure}[t]
\centering
\includegraphics[width=0.48\textwidth]{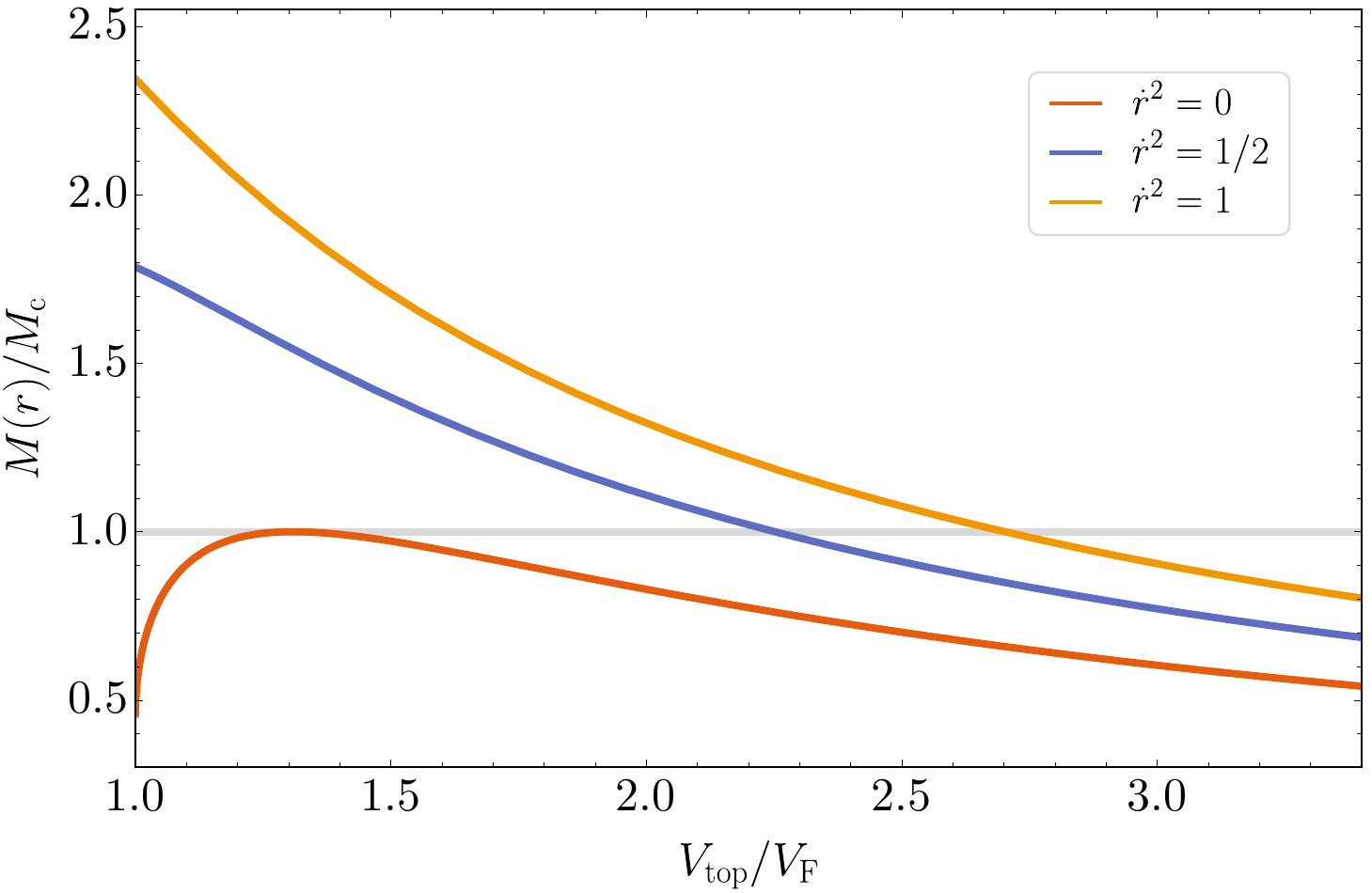}
\caption{Ratio of $M(r)$ to the critical bubble mass $M_\tn{c}$ as a function of the value of the potential at its maximum $V_\tn{top}$. For the parameters we pick $H_\tn{F}=2H_\tn{T}=3H_\sigma/2=M_\tn{Pl}$. Supercritical bubbles consistent with causality can arise from a generic potential, where the value at the top is an $\mathcal{O}(1)$ factor larger than the value in the false vacuum.}
\label{fig:HM}
\end{figure}

If we imagine that our Universes at higher vacuum energy are generated by a HM instanton, our first task is to compare the timescales for $\phi$ rolling down to its minimum and for the bubble wall evolution described in the previous Section.  In Appendix~\ref{app:HM} we find that the two are comparable. For simplicity, in the following we assume that the field rolls quickly to its minimum and then the bubble expansion starts.
If the hierarchy is reversed, our qualitative arguments are not affected. In this latter case, to generate a baby Universe we have to impose the conditions on the mass of the bubble $M > M_\tn{c}$ not only on Hubble in the false vacuum $H_\tn{F}$, but also to Hubble at the top of the potential $H_{\rm top}$. This does not change our discussion as $H_{\rm top}> H_\tn{F}$ and this implies $M(H_\tn{F}\to H_{\rm top})>M(H_\tn{F})$ as shown in Eq.~\eqref{eq:mass}. We can therefore impose our conditions on the critical mass using $H_\tn{F}$, which is relevant to the late-time evolution of the bubble, and they will be automatically satisfied also at early times. 

In Fig.~\ref{fig:HM} we show the mass of a typical HM bubble normalized to $M_\tn{c}$ for representative values of $H_\tn{F,\,T,\,top}$. If there is  an $\mathcal{O}(1)$ difference between $V_{\rm top}$ and $V_\tn{F}$, typical HM bubbles can be supercritical, if they are nucleated with $\dot r\neq 0$. Even when the typical bubble is too small to give birth to a new Universe, a bigger one can be nucleated either because of a quantum fluctuation of its size~\cite{PhysRevD.45.3469, Deng:2017uwc} or a second tunneling event~\cite{Farhi:1989yr, Fischler:1989se}. Most tunneling transitions do not create a new Universe in a higher energy vacuum either because the mass of the bubble is too small or because the tunneling transition goes to a vacuum of lower energy. However, an exponentially small fraction does, and this is all we need if the Multiverse is populated by eternal inflation. Lee and Weinberg tried to argue that the false vacuum occupies an exponentially small fraction of spacetime~\cite{Lee:1987qc}. However, these statements cannot be made precise in eternal inflation, as there is no way to define a gauge-invariant measure for observables throughout the Multiverse~\cite{Linde:1994gy,Vilenkin:1998kr,Guth:2007ng}. We do not try to estimate this volume fraction or the tunneling rates relevant to our processes because they are irrelevant to our purposes. Even if a single Universe with the observed value of the CC or $m_h^2$ is born, there are mechanisms that explain why we live there and not elsewhere in the Multiverse~\cite{Agrawal:1997gf, Arkani-Hamed:2004ymt, Geller:2018xvz, Giudice:2019iwl, Arkani-Hamed:2020yna, Strumia:2020bdy, Csaki:2020zqz, TitoDAgnolo:2021nhd, TitoDAgnolo:2021pjo, Giudice:2021viw, Bloch:2019bvc}, the best known being Weinberg's argument for the CC~\cite{Weinberg:1987dv}. 

In conclusion, we find no obstruction to the creation of new Universes from ``uptunneling" to higher energies, provided that enough minima in the landscape have positive vacuum energy during an inflationary phase. We have now written down a simple sector that provides eternal inflation (in Section~\ref{sec:eternal}) and studied the dynamics of false vacuum bubbles that lead to the generation of a Multiverse. As noted in Section~\ref{sec:eternal} our inflationary sector is in strong disagreement with measurements of the CMB, so there is still an unfinished task that lies ahead of us: explaining the last 60 $e$-folds of inflation.

%%%%%%%%%%%%%%%%%%%%%%%%%%%%%%%%%%%%%%%%%%%
\section{The Last Sixty \texorpdfstring{$\bm{e}$}{e}-Folds}\label{sec:CMB}
In this Section we slowly build up towards two models that can give the last sixty $e$-folds of inflation without violating our working hypotheses. We find it instructive to start from the simplest possible models, point out general tensions with the Swampland conjectures, and then extend them while trying to minimize the model-building complexity. This constructive approach shows that it is not so easy to evade both working hypotheses at the same time, even from a purely bottom-up perspective. It also highlights generic tensions between slow-roll inflation and the two hypotheses, both in the single-field and multifield cases.

To describe our Universe, compatibly with both working hypotheses, we imagine that one of the fields discussed in this Section is displaced from its minimum during eternal inflation and gives the last sixty $e$-folds of inflation in our patch.

The Section is organized as follows: we first show that it is the combination of the two hypotheses that excludes single-field slow-roll inflation, while it is easy to satisfy either one of the two. This suggests two possible ways to satisfy both: 1) We start with a flat potential and add spectator fields that take care of the refined dS conjecture, 2) We start with a steep potential and add new sources of friction. 

The three next subsections explore the first possibility. We begin by briefly recalling a tension between our working hypotheses and single-field slow-roll inflation, and then review a well-known model with a flat potential that can support many $e$-folds of inflation for small field excursions. Starting from this model we try to add a spectator field that dominates $|\nabla V|$, but gives a subdominant contribution to $V$, and find a general obstruction to this kind of ideas. However, more elaborate models of multifield inflation that satisfy both hypotheses exist~\cite{Aragam:2019omo, Kehagias:2018uem, Bravo:2020wdr}, but those that require large turning rates have other obstructions to UV completion in string theory~\cite{Aragam:2021scu}. We briefly comment on them in the following. 

We conclude the Section with an example in the second category, i.e.\ how a steep potential with friction from particle production can satisfy both working hypotheses. However, in this latter case other obstructions to UV-completing the model into string theory exist.

%%%%%%%%%%%%%%%%%%%%%%%%%%%%%%%%%%%%%%%%%%%
\subsection{Slow-Roll Single-Field Inflation}
We find it useful to review a generic tension between slow-roll single-field inflation and the combination of distance and refined dS conjectures that was already pointed out in~\cite{Achucarro:2018vey, Kehagias:2018uem}. 
If we assume slow-roll, the inflaton motion is described by a simple equation:
\be
\left|\frac{d\phi}{dN_e}\right| = \sqrt{2\epsilon} M_{\rm Pl} \simeq \sqrt{2\epsilon_V} M_{\rm Pl}\, .
\ee
The refined dS conjecture gives us a lower bound on $\epsilon_V$, summarized in our Eq.~\eqref{eq:WP2}. The bound implies\footnote{If we do not want to exponentially tune to be near a maximum, as discussed around Eq.~\eqref{eq:slowtuning}.}
\be
\Delta \phi \geq c N_e M_{\rm Pl}\, .
\ee
Therefore to have single-field slow-roll inflation, either we take
\be
c \leq \frac{1}{N_e^{\rm CMB}}\simeq \frac{1}{60}\, ,
\ee
which violates (at least in spirit) the refined dS conjecture, or we take $\Delta \phi$ super-Planckian, violating the distance conjecture. The physics behind these inequalities is very simple: if you have a steep potential, you need $\Delta \phi \gg M_{\rm Pl}$ for a long period of inflation. Otherwise, if you are determined to take $\Delta \phi \lesssim M_{\rm Pl}$, you need a flat potential $V^\prime/V \lesssim \frac{1}{N_e M_{\rm Pl}}$ to support $N_e$ $e$-folds of inflation. 

%%%%%%%%%%%%%%%%%%%%%%%%%%%%%%%%%%%%%%%%%%%
\subsection{Relaxing the Refined dS Conjecture}
\label{sec:hybrid}
The simple argument in the previous Section suggests that it is not particularly hard to satisfy just one of the two conjectures. In this Section we briefly recall one well-known model of inflation where one can inflate for a large number of $e$-folds with $\Delta \phi \ll M_{\rm Pl}$. Consider the potential for hybrid inflation~\cite{Linde:1993cn}
\be
V=\frac{1}{\lambda}(M^2-\lambda \sigma^2)^2+\frac{m^2}{2}\phi^2+\frac{g^2}{2}\phi^2\sigma^2\, . \label{eq:hybrid}
\ee
When $\phi > 2M/g$ the field $\sigma$ has a minimum at $\sigma=0$. If $M\gg m$, we can integrate out $\sigma$ and obtain an effective potential for the inflaton $\phi$
\be
V_{\rm eff} = \frac{M^4}{\lambda}+\frac{m^2}{2}\phi^2\, .
\ee
After some time, the inflaton crosses the critical value $\phi_\tn{c}=2M/g$ and the waterfall field $\sigma$ develops a new minimum. This ends inflation abruptly, with both fields settling into the new global minimum~\cite{Linde:1993cn}.

From Eq.~\eqref{eq:hybrid}, the number of $e$-folds is
\beq
\begin{aligned}
N_e(\Delta \phi)&=\int^{2M/g+\Delta \phi}_{2M/g}\frac{d\phi}{M_{\rm Pl}\sqrt{2\epsilon_V}} \\
&=\frac{\Delta \phi^2}{4 M_{\rm Pl}^2}+\frac{\Delta \phi M}{g M_{\rm Pl}^2}+\frac{M^4}{\lambda m^2 M_{\rm Pl}^2}\log\left(1+\frac{g\Delta \phi}{2M}\right)\, .  \label{eq:efolds}
\end{aligned}
\eeq
If we make the technically natural choice $g\ll1$, then $\Delta \phi/\phi_\tn{c} \ll 1$ and we can easily invert the previous equation to get the slow-roll result
\be
\Delta \phi \simeq M_{\rm Pl} N_e \sqrt{2 \epsilon_V} \simeq \frac{\lambda m^2 M_{\rm Pl}^2}{g M^3} N_e\, .
\ee
This shows in an explicit model that we can easily satisfy the distance conjecture if the inflaton potential is sufficiently flat
\be
\sqrt{\epsilon_V} \lesssim \frac{1}{N_e}\, .
\ee
Similar considerations apply to different models with flat potentials. Another simple example that displays the same properties are $\alpha$-attractors~\cite{Kallosh:2013yoa, Roest:2015qya, Galante:2014ifa} that were discussed in the context of the distance conjecture in~\cite{Scalisi:2018eaz}.
We do not give more details on hybrid inflation or other models here, as they have already been extensively studied in the literature.

%%%%%%%%%%%%%%%%%%%%%%%%%%%%%%%%%%%%%%%%%%%
\subsection{Multifield Inflation}
The simplest option for the last 60 $e$-folds is to take models similar to those in the previous Section and add spectator fields that make the potential steeper in directions orthogonal to the inflationary trajectory. This is possible, but not completely trivial to do, as we can see by considering the simplest possible setup. We add to our favorite inflationary model a new decoupled field $\phi_\tn{C}$. The new potential for inflation looks like
\beq
V(\phi, \phi_\tn{C})=V_\tn{I}(\phi)+V_\tn{C}(\phi_\tn{C})\, ,
\eeq
with $V_\tn{I}$ the hybrid inflationary potential in the previous Section. To drive inflation from $V_\tn{I}$ we need
\be
V_\tn{I} \gg V_\tn{C}
\ee
throughout the field range spanned by $\phi$ during inflation. In the same range we also need
\be
\frac{d V_\tn{I}}{d\phi} \lesssim \frac{V_\tn{I}}{M_{\rm Pl}} \lesssim \frac{dV_\tn{C}}{d\phi_\tn{C}}\, .
\ee
This second set of inequalities allows to satisfy the refined dS conjecture
\be
|\nabla V| \simeq \frac{dV_\tn{C}}{d\phi_\tn{C}} \gtrsim \frac{V_\tn{I}}{M_{\rm Pl}}\, .
\ee
This model has the same general problem of slow-roll single-field inflation for any $V_\tn{I}$, since the problem is purely in the $\phi_\tn{C}$ direction. The above equation can be rewritten as
\be
\frac{d V_\tn{C}}{d\phi_\tn{C}} \geq c H^2 M_{\rm Pl}\, , \label{eq:conj1}
\ee
so the typical distance traveled by $\phi_\tn{C}$ during inflation is
\be
\Delta \phi_\tn{C} \simeq (H\Delta t) \frac{1}{H^2} \frac{d V_\tn{C}}{d\phi_\tn{C}} \geq c N_e M_{\rm Pl}\, .
\ee
The way around this general tension is to make some directions in field space compact. In this case the long path of $\phi_\tn{C}$ can correspond to a small field excursion. Explicit models consistent with both of our working hypotheses have been constructed~\cite{Aragam:2019omo}. 

The potentials considered in~\cite{Aragam:2019omo} are a bottom-up construction and it would be worth it to check if they can be embedded in string theory. However, this general idea based on the geometry of the inflaton trajectory can potentially provide the last 60 $e$-folds. In the next Section we give another example of an inflationary sector consistent with both our working hypotheses.
 
%%%%%%%%%%%%%%%%%%%%%%%%%%%%%%%%%%%%%%%%%%%
\subsection{Warm Inflation}
We consider the model presented in~\cite{Berghaus:2019whh}, where the inflaton has the following potential:
\be
V(\phi, \sigma)=V_\tn{I}(\phi, \sigma)+\frac{\alpha}{16\pi}\frac{\phi}{f}\Tr\left[F\widetilde F\right]\, .
\ee
$V_\tn{I}$ is once again the hybrid inflation potential in Eq.~\eqref{eq:hybrid} and $F$ is the field strength of a new confining gauge group, with gauge coupling $\alpha$. As $\phi$ rolls down its potential, it can copiously produce gauge bosons that then thermalize, producing a radiation bath. The thermal bath slows down the inflaton via a friction term,
\beq
\begin{gathered}
\ddot \phi + 3 H(1+Q) \dot \phi + V^\prime=0\, , \\
Q \equiv \frac{\Upsilon(T)}{3H} \simeq \frac{\kappa \alpha^5 T^3}{3 H f^2}\, .
\label{eq:Q}
\end{gathered}
\eeq
The numerical coefficient $\kappa$ is $\mathcal{O}(100)$ and can be obtained on the lattice~\cite{Moore:2010jd} as the parametric dependence of $Q$ on $T, \alpha$ and $f$, see also~\cite{Laine:2016hma} for a review. For an $\tn{SU}(N)$ confining gauge group, $\kappa$ is well approximated by $N^5$.

We imagine to be in a regime where $H, T \gg \Lambda_g$, with $\Lambda_g$ the confining scale of the new gauge group, so that the coupling to $F\widetilde F$ does not modify the inflaton potential. In~\cite{Berghaus:2019whh} it was verified that this assumption is consistent with the values of the parameters that we consider in the following. We further imagine that the inflaton potential dominates the energy budget of the Universe:
\be
H^2 = \frac{1}{3M_{\rm Pl}^2}\left(V+\frac{\dot \phi^2}{2} +\rho_R\right) \simeq \frac{V}{M_{\rm Pl}^2}\, .
\ee
This is not hard to achieve if $\phi$ is slow-rolling. In this regime $\dot \phi^2 \ll V$ and
\beq
\begin{aligned}
\dot \phi &\simeq \frac{V^\prime}{3 H (1+Q)}\, , \\
\rho_R &\simeq Q \dot\phi^2\, . \label{eq:temperature}
\end{aligned}
\eeq
If $Q\gg 1$ the main source of friction is the thermal bath rather than Hubble expansion and we can have a large number of $e$-folds on a steep potential. To see this, it is useful to define the two quantities
\be
\xi \equiv \frac{V}{M_{\rm Pl} V^\prime}\, , \quad Q_\tn{S} \equiv  \left[\left(\frac{\kappa \alpha^5}{f^2}\right)^4\frac{M_{\rm Pl}^4 V}{g_*^3}\right]^{1/7}\, .
\ee
We can then use Eqs.~\eqref{eq:Q} and~\eqref{eq:temperature} to write $Q$ as
\be
Q(\xi)= \xi^{-6/7}Q_\tn{S}\, ,
\ee
and show that the number of $e$-folds is maximized at large values of $\xi$,
\beq
\begin{aligned}
N_e&=\int \frac{d \phi}{M_{\rm Pl}} \xi(1+\xi^{-6/7}Q_\tn{S}) \\
&\simeq \int \frac{d \phi}{M_{\rm Pl}} \xi^{1/7}Q_\tn{S}\, .
\end{aligned}
\eeq
So the number of $e$-folds is saturated at the largest $\xi$ compatible with the de Sitter conjecture, i.e.\ $\xi=1/c$. We want to know if $N_e \gtrsim 60$ is compatible with $\xi=1/c\simeq 1$ and observational constraints on the model. 

In the strong regime the scalar power spectrum is dominated by temperature fluctuations~\cite{Graham:2009bf,Mirbabayi:2022cbt,Berghaus:2024zfg}
\be
\Delta_R^2 \simeq\frac{1}{4\pi^2}\frac{H^3T}{\dot \phi^2}\widetilde{F}_\tn{M}(Q)\, ,
\label{eq:DR}
\ee
where
\be
\begin{aligned}
\widetilde{F}_\tn{M}(Q)&\equiv 5.4\times 10^{-5}Q^7 \\
&\phantom{{}={}}+168Q\left[\frac{1}{3}\left(1+\frac{9Q^2}{25}\right)+\frac{2}{3}\tanh\left(\frac{1}{30Q}\right)\right]\, .
\end{aligned}
\ee
The tensor power spectrum is the usual one $\Delta_h^2 \simeq \frac{2}{\pi^2}\frac{H^2}{M_{\rm Pl}^2}$, and for large $Q$ $n_s-1$ in this model reads~\cite{Berghaus:2024zfg}
\be
n_s -1\simeq 8\epsilon_Q-6 \eta_Q\, ,
\ee
where we defined $\epsilon_Q \equiv \epsilon_V/(1+Q)$ and the same for $\eta_Q$. In our low-scale inflation model $\eta_Q \gg \epsilon_Q$ and the measurement of $n_s-1$~\cite{Planck:2018jri} effectively fixes $\eta_Q \simeq 6\times 10^{-3}$.

To assess the compatibility of this construction with experiment and $N_e \gtrsim 60$ ($Q_\tn{S} \gtrsim 10^2$) it is useful to invert the relation between observables and parameters of the model for $Q\gg 1$,
\beq
\begin{aligned}
\phi_\tn{c} &\simeq \frac{M_{\rm Pl}}{\xi Q(\xi) \eta_Q}\, , \\
f&\simeq M_{\rm Pl}\frac{(\Delta_R^2)^{1/6}\kappa^{1/2}\alpha^{5/2}}{\widetilde{F}_\tn{M}(Q)^{1/6}g_*^{1/3}Q_\tn{S}^{7/6}}\, ,  \\
H&\simeq M_{\rm Pl} \frac{(\Delta_R^2)^{2/3}g_*^{1/6}}{\widetilde{F}_\tn{M}(Q)^{2/3}Q_\tn{S}^{7/6}}\, .
\end{aligned}
\eeq
Recall from Section~\ref{sec:hybrid} on hybrid inflation that $\phi_\tn{c}$ roughly corresponds to the maximal field excursion in the model, so we want $\phi_\tn{c} \lesssim M_{\rm Pl}$. If we consider $\alpha = 0.1$ and $g_* = 10$ that are easy to UV-complete, and then take $\xi=1$, $Q_\tn{S} = 300$ and $\Delta_R^2 \simeq 10^{-9}$, we obtain
\be
H \simeq 300\;{\rm GeV}\, ,\quad f\simeq 10^{-8}\, M_{\rm Pl}\, , \quad \phi_\tn{c} \simeq \frac{5 M_{\rm Pl}}{9}\, .
\ee
The bound on the tensor-to-scalar ratio is automatically satisfied, as it just requires $H \lesssim 10^{-5} M_{\rm Pl}$.
Increasing the number of $e$-folds decreases $\phi_\tn{c}$, but also $H$.

Hence, it is possible to have a warm inflation model which is consistent with our working hypotheses, and gives rise to 60 $e$-folds of inflation consistently with observations. 
However, we see immediately that there is a model-building price to pay. The inflaton behaves at the same time as a compact field (in the coupling to $F\widetilde F$) and a non-compact  field (in $V_\tn{I}$). This is a consequence of requiring $f\ll \phi_\tn{c}$ and makes UV-completing this model very hard in the context of string theory, as was already discussed extensively in~\cite{McAllister:2016vzi} for the relaxion~\cite{Graham:2015cka}.

It is legitimate to wonder if changing the coupling of $\phi$ to the thermal bath one could find models that are easier to UV-complete or that support a larger number of $e$-folds. There is a generic obstruction to this possibility. The friction term $\Upsilon(T)$ can be computed from the two-point function of $\phi$ in a thermal bath~\cite{Laine:2016hma}. The same two-point function gives a large correction to the $\phi$ potential making it steep enough to nullify the effects of thermal friction~\cite{Yokoyama:1998ju,Berghaus:2024zfg}. Consider for example a Yukawa coupling to a fermion of mass $m$:
\be
V \supset y \phi \bar \psi \psi \, .
\ee
The friction term at one-loop is given by
\be
 \Upsilon_\tn{Y} = \frac{y^2}{8 \pi}\frac{(m_{\phi, \text{eff}}^2 - 4m^2)^{\frac{3}{2}}}{m_{\phi, \text{eff}}^2}  \left(\frac{e^{m_{\phi, \text{eff}}/2T}-1}{e^{m_{\phi, \text{eff}}/2T}+1} \right)\, , \label{eq:Yfriction}
 \ee
 where $m_{\phi, \text{eff}}^2$ is the one-loop effective mass of $\phi$ that includes thermal and non-thermal radiative corrections.
Thermal effects become important and can dominate the friction term if $yT \gg m_\phi, m$. In this limit $ \Upsilon_\tn{Y}(T) \sim T$, the $T^2/f^2$ suppression of the previous model is not present and one could hope to get out many more $e$-folds out of a generic steep potential. 

Note that in this regime one has to be careful about the validity of the one-loop approximation for $\Upsilon_\tn{Y}$, that holds only if 
\be
y \lesssim \frac{\min[m, m_\phi]}{y T} \lesssim 1\, .
\ee
If this condition is not satisfied Eq.~\eqref{eq:Yfriction} changes and in some limits one cannot compute explicitly $\Upsilon_\tn{Y}$, but our considerations below still hold at the qualitative level.

The problem is that the Yukawa coupling gives both a thermal and a non-thermal correction to the mass of $\phi$. If we imagine new particles coming in at some scale $M_{\rm UV}$, then the validity of our EFT requires $M_{\rm UV} > T$. The same limit that makes thermal friction in Eq.~\eqref{eq:Yfriction} important ($yT \gg m_\phi, m$), makes the non-thermal correction to $m_\phi$ even more important: $m_{\phi, \text{eff}}^2 \simeq \delta m_\phi^2(T=0)\simeq y^2 M_{\rm UV}^2/16\pi^2$. In turn this makes $\eta_Q$ too large to support 60 $e$-folds of inflation,
\be
\eta_Q \simeq \frac{M_{\rm Pl}}{y M_{\rm UV}}\, ,
\ee
if we take $y$ perturbative and $M_{\rm UV}\lesssim M_{\rm Pl}$. This is just an example, but it illustrates a general problem of warm inflation that has to be circumvented by clever model-building as was done in the model~\cite{Berghaus:2019whh} that we considered. There the coupling $\phi F\widetilde F$ can be written in terms of the derivative of a current $\phi \partial_\mu K^\mu$ and integration by parts shows that the $\phi$ potential does not receive any correction. For other attempts along these lines see for example~\cite{Berera:1998px,Bastero-Gil:2016qru,Bastero-Gil:2019gao}. For more on the relationship between warm inflation and the Swampland, see instead~\cite{Motaharfar:2018zyb,Das:2018rpg,Brandenberger:2020oav,Das:2020xmh}.

%%%%%%%%%%%%%%%%%%%%%%%%%%%%%%%%%%%%%%%%%%%
\section{How Many \texorpdfstring{$\bm{e}$}{e}-Folds for a Multiverse?}\label{sec:tunneling}
In this Section we relax one of the two main assumptions made throughout the paper and we allow for the existence of dS minima. We do so to discuss another rather unexpected result. We find that the minimum number of $e$-folds needed to populate such an ``ordinary" Multiverse, i.e.\ one where the CC can take any value, can be close to $N_e \sim 60$.

The Multiverse is normally considered in the context of eternal inflation~\cite{Vilenkin:1983xq, Winitzki:2008zz, Guth:2007ng, Linde:1986fc, Linde:1986fd} where corners of spacetime inflate forever. We have already shown that in this context it is possible to populate a Multiverse without violating our two working hypotheses.

However, it is also easy to show, at least in toy models of the landscape, that we can populate a Multiverse large enough to explain the value of the CC (and simultaneously of the weak scale) with a finite number of $e$-folds and without eternal inflation. In this case we do not construct an explicit model of inflation that can realize the Multiverse, we just want to challenge the widespread belief that Multiverse $=$ eternal inflation. 

Note again that in this Section we assume that the landscape contains also dS minima and drop our working hypothesis in Eq.~\eqref{eq:WP2}. 

%%%%%%%%%%%%%%%%%%%%%%%%%%%%%%%%%%%%%%%%%%%%%%
\subsection{The Minimum Number of \texorpdfstring{$\bm{e}$}{e}-Folds}
The semiclassical rate of tunneling between two vacua can be schematically written as~\cite{Arkani-Hamed:2005zuc}
\be
\Gamma \simeq V M_*^4 e^{-S_\tn{E}}\, , \label{eq:rate}
\ee
where $S_\tn{E}$ is the bounce action that connects the two vacua~\cite{Coleman:1977py, Callan:1977pt, Coleman:1980aw} and $V$ is the volume of the causally connected region where the bubble of true vacuum can be nucleated. For simplicity, we take the landscape to be characterized by a single mass scale $M_*$. We can obtain an upper bound on $e^{-S_\tn{E}}$ from the existence of our Universe,
\be
e^{-S_\tn{E}} \lesssim \frac{1}{\tau_\tn{U} V_\tn{U} M_*^4}\, ,
\label{eq:Universe}
\ee
where $\tau_\tn{U}$ and $V_\tn{U}$ are the lifetime and volume of the observable Universe. For the volume $V$ in Eq.~\eqref{eq:rate} we consider an inflationary Universe of initial size $V_i \sim H_\tn{I}^{-3}$, where $H_\tn{I}$ is Hubble during inflation. After $N_e$ $e$-folds it has expanded to a size
$V(N_e)=V_i e^{3 N_e}$.

If a time $\Delta t= N_e H^{-1}_\tn{I}$ has elapsed since the beginning of inflation and we have a vacuum decay rate independent of time, the nucleation of bubbles is a Poisson process with mean $\Gamma \Delta t$. Therefore the expected number of bubbles $\langle N_\tn{b}\rangle $ nucleated after $N_e$ $e$-folds is 
\beq
\begin{aligned}
\langle N_\tn{b} \rangle &= \frac{M_*^4 e^{-S_\tn{E}}}{H_\tn{I}^4} e^{3 N_e} N_e \\
&\lesssim \frac{1}{H_\tn{I}^4 \tau_\tn{U} V_\tn{U}} e^{3 N_e} N_e \, .
\label{eq:bubbles}
\end{aligned}
\eeq
To scan the CC down from the Planck scale we need about $10^{120}$ bubbles\footnote{Assuming a uniform distribution of vacua around the small observed value of the CC~\cite{Arkani-Hamed:2005zuc}. See also the next Section.}  corresponding to 
\be
N_e\gtrsim 270+ \log\left(\frac{H_\tn{I}}{10^{16}\;{\rm GeV}}\right)^{4/3} 
\ee
$e$-folds of inflation. If we want to scan down also the weak scale we need $\simeq 10^{154}$ bubbles and at least 300  $e$-folds. This estimate is conservative, as we have neglected that some bubbles continue to inflate during this time, generating an exponentially large volume where new bubbles can form. We have counted only bubbles nucleated within the original inflating Universe. This allows us to avoid specifying UV details of the landscape (i.e.\ the probability distribution of bubbles that remain in a dS phase for some time), but still get a number of $e$-folds that guarantees enough bubbles to scan the CC\footnote{The word ``conservative" here might generate some confusion. Our lower bound is conservative because it ensures that we generate a big enough Multiverse in any landscape. However it is not conservative because in certain landscapes one needs much fewer $e$-folds.}. 

To turn this lower bound into an actual estimate we can introduce a simple toy model of the landscape.

%%%%%%%%%%%%%%%%%%%%%%%%%%%%%%%%%%%%%%%%%%%%%%
\subsection{A Better Estimate}
What we really care about is the tunneling rate from a typical point in the landscape where $\Lambda^4 \simeq M_{\rm Pl}^4$ to our small CC vacuum. We need enough $e$-folds to make the probability of this transition $\mathcal{O}(1)$. To be completely explicit, in our model the CC is a sum of many contributions 
\be
\Lambda_{\rm CC}=\sum_{i=1}^N M_i^4\, , \quad N \gg 1\, ,
\ee
with the dominant ones being $\mathcal{O}(M_{\rm Pl}^4)$. Some of these contributions come from the vacuum energy of scalar fields.
We imagine to be in a situation where a single tunneling event can take us to a vacuum where
\beq
\begin{aligned}
M_n^4 &= - \sum_{i\neq n} M_i^4 + \delta \Lambda_{\rm obs}\, , \\
 \delta \Lambda_{\rm obs} &\simeq {\rm meV}^4\, ,
\end{aligned}
\eeq
i.e we need a single jump of $\mathcal{O}(M_{\rm Pl}^4)$ in vacuum energy to reach our vacuum from a generic one with $|\Lambda_{\rm CC}| \simeq M_{\rm Pl}^4$. This would happen, for instance, in landscapes such as those of the Bousso-Polchinski scenario~\cite{Bousso:2000xa}. Of course, this does not mean that we can reach our vacuum from anywhere in the landscape with a single jump. In general we need many nucleation events. How many exactly? To estimate this number we need a (statistical) picture of the landscape. Imagine a vacuum distribution that scans finely the CC around zero:
\be
p(\Lambda)=\frac{1}{\sqrt{2\pi} \sigma_{\Lambda}}e^{-\frac{\Lambda^2}{2\sigma_{\Lambda}^2}}\, .
\ee
This is realized for instance by a ``friendly'' supersymmetric landscape~\cite{Arkani-Hamed:2005zuc}. If the landscape is generated by heavy fields with typical mass scales $M_* \simeq M_{\rm Pl}$ from dimensional analysis (or a more explicit calculation~\cite{Arkani-Hamed:2005zuc}) we expect $\sigma_{\Lambda} \simeq n M_{\rm Pl}^4$, where $n$ is a number that depends on the structure of the landscape and can be larger than one. Most vacua have $|\Lambda_{\rm CC}|\simeq n M_{\rm Pl}^4$ and a typical tunneling event takes us from a vacuum with large CC to another vacuum with large CC. Landing by chance in a window with width $\Delta \Lambda \sim \delta \Lambda_{\rm obs} $, requires an exponentially large amount of luck
\beq
\begin{aligned}
P[\Lambda \leq \Lambda_{\rm CC}\leq \Lambda+\delta \Lambda_{\rm obs}]&= \int_{\Lambda}^{\Lambda+\Delta \Lambda} p(\Lambda_{\rm CC}) d\Lambda_{\rm CC} \\
&\simeq \frac{\delta \Lambda_{\rm obs}}{\sigma_{\Lambda}} \simeq 10^{-120}\, . \label{eq:CCp}
\end{aligned}
\eeq
So we need about $10^{120}$ tunneling events to have an $\mathcal{O}(1)$ probability to end in our vacuum at least once, which corresponds to solving $\langle N_\tn{b} \rangle(N_e) \simeq 10^{120}$.

To estimate the probability of nucleating enough bubbles, for concreteness we take the simple potential in Coleman's papers on semiclassical vacuum decay~\cite{Coleman:1977py, Coleman:1980aw} for the scalar that needs to tunnel, and add a generic CC:
\be
V_0(\phi)={\bar \Lambda}+\frac{\lambda}{8}\left(\phi^2-\frac{\mu^2}{\lambda}\right)^2\, ,
\ee
with $\lambda\simeq 1$, $\bar \Lambda\simeq M_{\rm Pl}^4$ and $\mu\simeq M_{\rm Pl}$. This potential has two minima $V_0(\phi_\pm)={\bar \Lambda}$ and one maximum $V_0(0)={\bar \Lambda}+\mu^4/8\lambda$. We add an unspecified perturbation to this potential such that the minima are split:
\be
V(\phi)=V_0(\phi)+V_1(\phi)\, .
\ee
We do not write down $V_1(\phi)$, but just parametrize the difference in vacuum energy between minima (that we call $\Delta V$) in terms of a dimensionless number $\omega$ and $V_0$,
\begin{equation}
\Delta V\equiv V_1(\phi_+)-V_1(\phi_-)= \omega \left(V_0(0)-{\bar \Lambda}\right) = \omega\frac{\mu^4}{8\lambda}\simeq \omega M_{\rm Pl}^4\, .
\end{equation}
This parametrization allows us to specify $\Delta V$ (or $\omega$) independently of the value of $\bar \Lambda$.
For simplicity we consider values of $\omega$ that allow for a thin-wall approximation of the decay rate. As we show in a few lines, this requirement is not very restrictive, it is actually looser than the requirement $\omega \lesssim 1$. For the discussion in this Section, we also define $\bar\Lambda\equiv CM_\text{Pl}^4$ and $\kappa\equiv M_\text{Pl}^{-2}$ for later convenience.

When tunneling between general minima, we have\footnote{Note that for simplicity we always take $\mu=M_{\rm Pl}$.}
\beq
\begin{aligned}
\bar \rho&=\frac{\bar \rho_0}{\left(\left(1+\frac{\bar \rho_0^2}{4R^2}\right)^2+\frac{\bar \rho_0^2}{L^2}\right)^{1/2}}\, , \\
\bar \rho_0 &= \frac{3}{\Delta V}\int_{\phi_-}^{\phi_+} d\phi \sqrt{2(V_0(\phi)-V_0(\phi_+))}= \frac{16}{\omega \mu}\, , \\
R^2 &=\frac{3M_\text{Pl}^2}{\Delta V}\, , \qquad L^2 =\frac{3M_\text{Pl}^2}{\bar \Lambda}\, . \label{eq:dStunnel}
\end{aligned}
\eeq
The thin-wall approximation for the tunneling rate is valid when~\cite{Coleman:1977py, Coleman:1980aw,Parke:1982pm}
\be
{\bar \rho}^2, \, R^2, \, L^2 \gg \frac{1}{\mu^2}\, .
\ee

%%%%%%%%%%%%%%%%%%%%%%%%%%%
\begin{figure}[!t]
\centering
\includegraphics[width=0.48\textwidth]{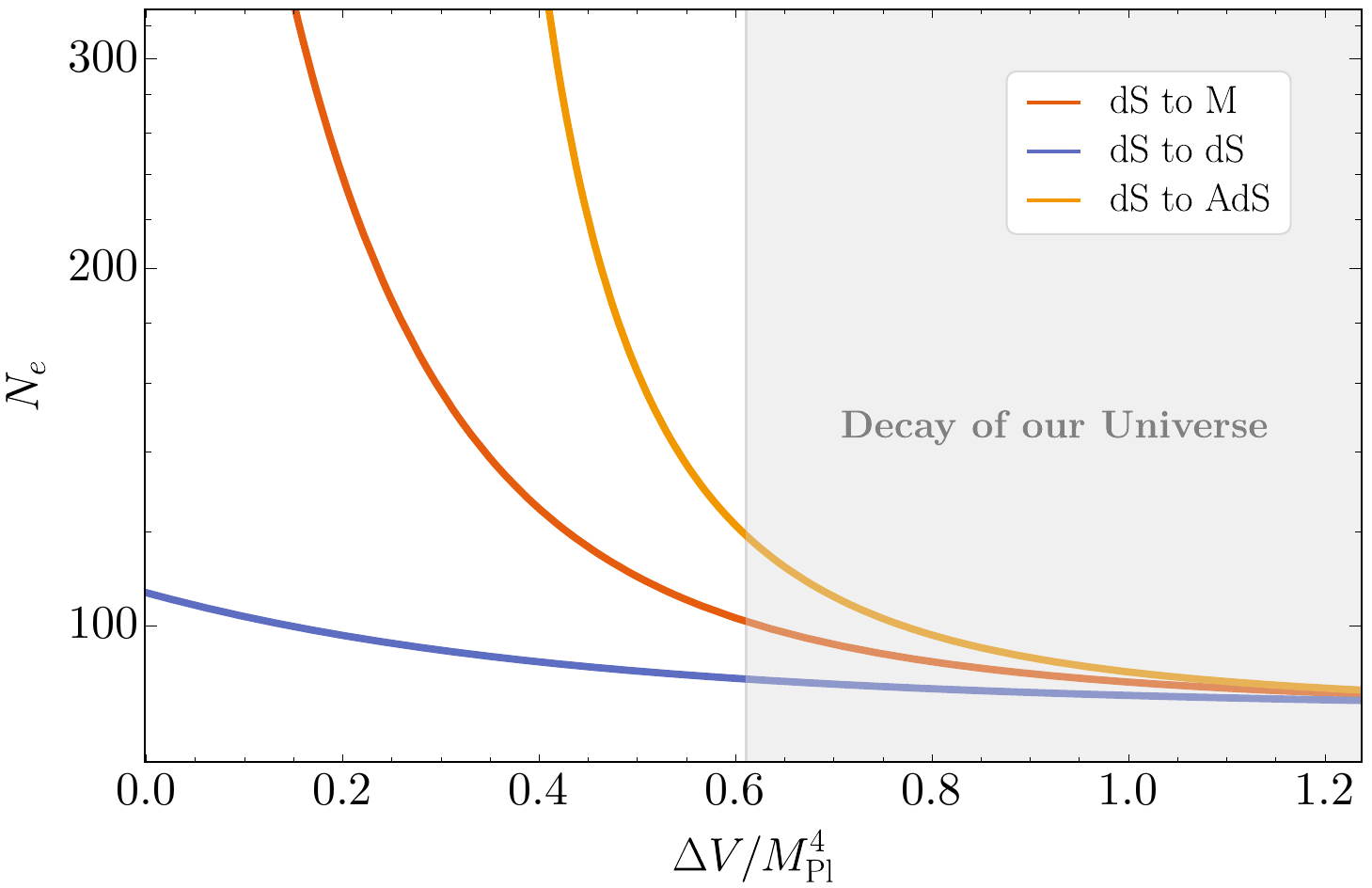}
\caption{Comparison of the minimum number of $e$-folds to explain the CC. In the shaded region our Universe decays too rapidly.}
\label{fig:decay}
\end{figure}
%%%%%%%%%%%%%%%%%%%%%%%%%%%%

For $\lambda =1$, these roughly translate to $C\ll3, \omega\ll10$, and the thin wall approximation holds. In this limit the tunneling rate $\Gamma$ has a simple analytical expression, $\Gamma/V\simeq M_\text{Pl}^4e^{-B}$, with\footnote{In the appropriate limits our expressions agree with the classical results in~\cite{Coleman:1977py, Coleman:1980aw}. In the absence of gravity, $\kappa\to 0$, we get $B=27 \pi^2 \sigma^4/(2 \Delta V^3)$. For a dS to Minkowski transition ($\bar \Lambda\to 0$) we reproduce the result in (3.16) of~\cite{Coleman:1980aw} and for a Minkowski to AdS transition ($C\to -\omega/8$) we reproduce (3.19) of the same paper. Our expression for $B$ also agrees with~\cite{Parke:1982pm}.}
\begin{equation}
\begin{aligned}
B&\simeq\frac{12 \pi ^2 }{\kappa ^2 {\bar \Lambda}  (\Delta V+{\bar \Lambda} ) \sqrt{\left(4 \Delta V+3 \kappa  \sigma ^2\right)^2+48 \kappa  {\bar \Lambda}  \sigma ^2}} \\
&\phantom{{}={}}\times\biggl[6 \kappa {\bar \Lambda}  \sigma^2+\Delta V\biggl(4 \Delta V+3\kappa  \sigma ^2 \\
&\phantom{{}={}}- \sqrt{\left(4 \Delta V+3 \kappa  \sigma ^2\right)^2+48 \kappa  {\bar \Lambda}  \sigma ^2}\,\biggr)\biggr]\, ,
\label{eq:rateCD}
\end{aligned}
\end{equation}
where $\sigma = \bar \rho_0 \Delta V/3$ is the bubble wall tension. In our case
\beq
\begin{aligned}
B&\simeq\frac{12 \pi ^2 }{C (8 C+\omega )\sqrt{768 C+(8+3 \omega)^2}} \\
&\phantom{{}={}}\times\left[128 C+\omega  \left(8+3 \omega-\sqrt{768 C+(8+3 \omega)^2}\,\right)\right]\, .
\end{aligned}
\eeq
We can already see that the minimum number of $e$-folds is very UV-dependent, in the sense that it will change polynomially when we change $\mathcal{O}(1)$ numbers in the potential such as $\lambda$ or $\omega$. Now we can repeat the estimates in the previous Section requiring $10^{120}$ bubbles, $\langle N_\tn{b}\rangle(N_e) \simeq 10^{120}$. This gives
\beq
\begin{aligned}
N_e &\simeq \frac{B}{3}+\frac{1}{3}\log\langle N_\tn{b}\rangle+\frac{4}{3}\log\frac{H_\tn{I}}{M_\text{Pl}} \\
&\simeq85+\frac{B}{3}+\frac{4}{3}\log\frac{H_\tn{I}}{10^{16} \GeV}\, . 
\label{eq:efolds1}
\end{aligned}
\eeq
For instance, for $\omega=C=1$ we have $N_e \simeq 100$. We can focus on three notable cases:
\begin{enumerate}
\item de Sitter to Minkowski, $C\rightarrow 0$
\item de Sitter to de Sitter, for example with $C=1$
\item de Sitter to Anti-de Sitter, for example with $C=-1/3$
\end{enumerate}
Comparing those three situations in Fig.~\ref{fig:decay} shows how tunneling from de Sitter to de Sitter can significantly reduce the minimum number of $e$-folds compared to the other cases. Once again, the existence of our Universe sets a lower bound on $B$, in analogy with Eq.~\eqref{eq:Universe}. In other words, tunneling into a deeper AdS minimum must be sufficiently suppressed, which translates into the condition $\omega\lesssim5$. Within the corresponding non-shaded area, our requirements can be satisfied with $\mathcal{O}(100)$ $e$-folds of inflation. 

A real Multiverse is populated by a combination of the transitions in our three case studies, but if enough minima in the landscape are uplifted to dS during inflation, the final result will not be too far from our blue line in Fig.~\ref{fig:decay}.

In addition to our main point on the number of $e$-folds, it is interesting to comment briefly on some limits of our three case studies. The dS to AdS transition becomes zero probability when $\omega$ becomes small, because $\bar \rho$ grows to infinity and gravity is stabilizing the false vacuum as for the Minkowski to AdS transitions considered in~\cite{Coleman:1980aw}. In the dS to Minkowski case the same happens for a different reason. When $\omega= 0$ we are looking at a Minkowski to Minkowski transition that is suppressed by the infinite volume of spacetime $\Gamma \sim e^{-a V}$. In the dS to dS case (blue line in Fig.~\ref{fig:decay}) we have a horizon which keeps $\Gamma$ finite. The tunneling problem is similar to quantum mechanics, where the probability remains finite for $\omega \to 0$.

Note also that some AdS to AdS transitions, naively described by our equations, are not physical. One can take $\omega=0$ and $C=-1/12$ and find the puzzling result $B\to -\infty$, i.e.\ an infinite transition rate. The same happens for $\omega=0$ and $C\to 0^-$. Furthermore, $B$ remains negative for all values of $C\in [-1/12, 0]$. However, it is easy to check that at the two singular points $\bar \rho \to \infty$ and in between $\bar \rho$ is larger than the AdS curvature radius.

%%%%%%%%%%%%%%%%%%%%%%%%%%%%%%%
\section{Conclusion}
We have begun this work with the construction of a simple model of eternal inflation on a steep potential without trans-Planckian field excursions. We have then shown that a Multiverse can be populated from tunneling transitions that move fields from a low-energy true vacuum to a high-energy false vacuum. These two ingredients allow to generate a Multiverse compatible with the distance and refined dS conjectures and the arguments in~\cite{Dvali:2013eja, Dvali:2014gua, Dvali:2017eba, Dvali:2020etd, Dvali:2021kxt}. 

It is interesting to notice that it is much harder, but possible, to write down models that give the last 60 $e$-folds of inflation and are compatible both with the two conjectures and with the CMB. In our discussion around the last 60 $e$-folds of inflation, in Section~\ref{sec:CMB}, we have also shown some generic tensions between the two conjectures and slow-roll inflation, both in the single-field and multifield cases. We have further studied the addition of a new source of friction in the form of a thermal bath and shown that albeit consistent with both conjectures, viable models of warm inflation present other obstructions to their UV-completion.

To conclude, we have relaxed one of our initial assumptions, the absence of dS minima in the landscape, to challenge a long-held belief in the community. The standard lore is that a Multiverse that explains the CC must come from eternal inflation. Our estimates indicate that a number of $e$-folds of $\mathcal{O}(100)$ might be sufficient.

%%%%%%%%%%%%%%%%%%%%%%%%%%%%%%%%%%%%%%%%%%%%%%%%%%%%%%%%%%%%%%%%%%%%%%%%%%%
%%%%%%%%%%%%%%%%%%%%%%%%%%%%%%%%%%%%%%%%%%%%%%%%%%%%%%%%%%%%%%%%%%%%%%%%%%%

%%%%%%%%%%%%%%%%%%%%%%%%%%%%%%%%%%%%%%%%%%%%%%%%%%%%%%%%%%%%%%%%%%%%%%%%%%%
%%%%%%%%%%%%%%%%%%%%%%%%%%%%%%%%%%%%%%%%%%%%%%%%%%%%%%%%%%%%%%%%%%%%%%%%%%%
\mySections{Acknowledgments}

We thank A.\ Bedroya, K.\ Berghaus, M.\ Geller, L.\ Santoni, R.\ Wald for useful discussions and comments on the manuscript. This research is supported by a grant under the France And Chicago Collaborating in The Sciences (FACCTS) program.  G.\ R.\ acknowledges funding from the European Union's Horizon 2020 research and innovation programme under the Marie Sk{\l}odowska-Curie actions Grant Agreement no 945298-ParisRegionFP.   The work of L.T.\ W.\ is also supported by DOE grant DE-SC-0013642.  

%%%%%%%%%%%%%%%%%%%%%%%%%%%%%%%%%%%%%%%%%%%%%%%%%%%%%%%%%%%%%%%%%%%%%%%%%%%
%%%%%%%%%%%%%%%%%%%%%%%%%%%%%%%%%%%%%%%%%%%%%%%%%%%%%%%%%%%%%%%%%%%%%%%%%%%

%\widetext
%\bigskip\bigskip
%\begin{center}
%\LARGE \bf Appendix
%\end{center}
%\bigskip
\twocolumngrid
\appendix
\section{Dynamics of False Vacuum Bubbles}\label{app:mass}
We start from the equation of motion in the main text:
\begin{align}
\beta_\tn{F}-\beta_\tn{T}&=4\pi G_\tn{N} \sigma r\, , \label{eq:masterA}\\
\beta_\tn{F}&=\pm\sqrt{1-H_\tn{F}^2 r^2+\dot r^2}\, ,\\
\beta_\tn{T}&=\pm\sqrt{1-H_\tn{T}^2r^2-\frac{2G_\tn{N} M}{r}+\dot r^2}\, .
\end{align}
It is useful to introduce a series of definitions that allow to put Eq.~\eqref{eq:masterA} in a form that is manifestly easy to solve. First of all we define the energy scales relevant to the problem as in Eq.~\eqref{eq:def}:
\be
H_\sigma \equiv 4\pi G_\tn{N} \sigma\, , \quad \Delta H^2_+\equiv H_\sigma^2+H_\tn{F}^2-H_\tn{T}^2\, .
\ee
Then we rescale the coordinates
\be
z^3\equiv \frac{\Delta H^2_+ r^3}{2 G_\tn{N} M}\, , \quad \tau^\prime = \frac{\Delta H^2_+}{2 H_\sigma}\tau\, ,
\ee
and finally we introduce three dimensionless parameters
\beq
\begin{aligned}
E(M)&\equiv - \frac{4 H_\sigma^2}{[(2 G_\tn{N} M)^2 \Delta H^8_+]^{1/3}}\, , \\
 \gamma &\equiv \frac{2 H_\sigma}{\Delta H_+}\, , \quad \alpha\equiv \frac{H_\tn{T}^2}{\Delta H^2_+}\, .
\end{aligned}
\eeq
After some algebra Eq.~\eqref{eq:masterA} reduces to
\beq
\begin{aligned}
\left(\frac{dz}{d\tau^\prime}\right)^2&=E-V(z) \label{eq:EOM}\, , \\
V(z)&= - \frac{(1-z^3)^2}{z^4}-\gamma^2 \left(\frac{1}{z}+\alpha z^2\right)\, .
\end{aligned}
\eeq
The formal solution to this equation is well known and takes the form
\be
\int^z \frac{dz^\prime}{\sqrt{E-V(z^\prime)}}=\tau^\prime\, .
\ee
A detailed analysis of the solutions was carried out in~\cite{Blau:1986cw} for $H_\tn{T}=\alpha=0$. Its qualitative features remain valid in our more general case. In particular $V(z)$ has a single maximum at
\be
z_M^3=\frac{-2 +\gamma^2+\sqrt{36-4\gamma^2+32\alpha \gamma^2+\gamma^4}}{4(1+\alpha \gamma^2)}\, ,
\ee
and one can define a critical mass for the bubble from
\be
E(M_\tn{c})=V(z_M)\, . \label{eq:Mc}
\ee
As discussed in the main text, for $M>M_\tn{c}$ we can generate a Universe from the false vacuum bubble. We use Eq.~\eqref{eq:Mc} to derive the expressions for the critical mass discussed in the main text. We find that $M_\tn{c}$ is the maximum of $M(r)$ when $\dot r=0$ and $\beta_\tn{F} > 0$ and write the compact expression in Eq.~\eqref{eq:Mcrit},
\begin{equation}
M_\tn{c}=\frac{R^3_\tn{c}\left(H^2_\tn{F}-H^2_\tn{T}-H^2_\sigma\right)}{2G_\tn{N}}+\frac{R_\tn{c}H^2_\sigma \left(6R^2_\tn{c}H^2_\tn{F}-4\right)}{3G_\tn{N}\left(H^2_\tn{F}-H^2_\tn{T}-H^2_\sigma\right)}\, ,
\end{equation}
by solving $dM/dr=0$. In general Eq.~\eqref{eq:EOM} has to be solved numerically, but we can gain some analytical insight by taking the limit $M\to \infty$. In this limit
\be
z^3(\tau^\prime)=1- e^{-3\tau^\prime}\, . \label{eq:Minfty}
\ee
This result means that we approach an infinite size for the bubble exponentially, because $z\to 1$ means $r \to\infty$. In obtaining the previous equation we considered the only possible initial condition, i.e.\ $z=0$. 

The effective Hubble rate for this exponential growth can be read from the definition of $\tau^\prime$ and it is
\be
H_{\rm eff}=\frac{\Delta H^2_+}{2 H_\sigma}=\frac{\Delta H_+}{\gamma}\, . \label{eq:Heff}
\ee
This solution is not physical because the bubble forms in dS space and contains dS space, so we have horizons of size $\sim 1/H_\tn{F,\,T}$.
For now we just note that Eqs.~\eqref{eq:Minfty} and Eq.~\eqref{eq:Heff}, albeit unphysical, are useful to identify the timescale of the bubble wall dynamics for an observer inside the bubble.

%%%%%%%%%%%%%%%%%%%%%%%%%%%%%%%%%%%%%%%%%%%%%%%%%%%%%%%%%%%%%%%%%%%%%
\section{Dynamics of Supercritical Hawking-Moss Bubbles}\label{app:HM}
We consider a scalar field HM bubble at the top of the potential barrier between two minima. In one $H_{\rm top}^{-1}$ time, the classical motion of our scalar $\phi_\tn{t}$ has changed its position by 
\be
\Delta \phi_\tn{t} \simeq \frac{V^\prime}{H_{\rm top}^2} \gtrsim \frac{V_\tn{F}}{M_{\rm Pl} H_{\rm top}^2} \simeq \frac{V_\tn{F}}{V_{\rm top}}M_{\rm Pl}\, .
\ee
This estimate assumes that the field is already slightly displaced from the top (where $V^\prime = 0$) shortly after the tunneling event, and that in the rest of the potential the first inequality in Eq.~\eqref{eq:WP2} holds.
From Eq.~\eqref{eq:WP1} $\Delta \phi_\tn{t}^{\rm max}\leq M_{\rm Pl}$, so it takes at most a time
\be
\Delta t_{\text{roll}} \simeq \frac{1}{H_{\rm top}} \frac{V_{\rm top}}{V_\tn{F}} = \frac{1}{H_\tn{F}}
\ee
to reach the minimum. We can compare this with the time it takes for the spacetime inside the bubble to expand from the point of view of the inside observer. For a sufficiently large mass (see the discussion around Eq.~\eqref{eq:Minfty}) the time it takes for the bubble radius to change by $\mathcal{O}(1)$ is
\be
\Delta t_{\rm inside} \simeq \frac{\gamma_{\rm wall}}{H_{\rm eff}}\, ,
\ee
where $\gamma_{\rm wall}$ is the Lorentz factor of the wall and $H_{\rm eff}$ is in Eq.~\eqref{eq:Heff}.
The outside observer sees the bubble wall expanding at most at the speed of light
\be
\Delta t_{\rm outside} \simeq \frac{\gamma_{\rm wall}}{H_{\rm top}}\, .
\ee
We took the two observers close enough to the wall to neglect differences in the metric with respect to the wall itself. So we obtained these two estimates by rescaling the proper time $\tau$ of the wall by the boost factor $\gamma_{\rm wall}$.

To conclude, we can estimate the boost of the wall by assuming that its kinetic energy comes entirely from the difference in vacuum energy between the two sides
\be
\gamma_{\text{wall}} \simeq \frac{R_\tn{HM} (V_{\rm top}-V_\tn{T})}{\sigma} \simeq  \frac{(V_{\rm top}-V_\tn{T})}{ H_{\rm top} \sigma} 
\ee 
and find that 
\be
\frac{\Delta t_{\rm inside}}{\Delta t_{\text{roll}}} \simeq \frac{H_\tn{F}}{H_{\rm top}}\frac{H_{\rm top}^2-H_\tn{T}^2}{\Delta H^2}\, .
\ee
The hierarchy between the two timescales depends on the details of the potential, and in general they are comparable.

%%%%%%%%%%%%%%%%%%%%%%%%%%%%%%%%%%%%%%%%%%%%%%%
%%%%%%%%%%%%%%%%%%%%%%%%%%%%%%%%%%%%%%%%%%%%%%%

\bibliographystyle{utphys}
\bibliography{refs}

\providecommand{\href}[2]{#2}\begingroup\raggedright\begin{thebibliography}{100}

\bibitem{Palti:2022edh}
E.~Palti, ``{The swampland and string theory},''
  \href{http://dx.doi.org/10.1080/00107514.2022.2103275}{{\em Contemp. Phys.}
  {\bf 62} (2022) no.~3, 165--179}.

\bibitem{Ooguri:2006in}
H.~Ooguri and C.~Vafa, ``{On the Geometry of the String Landscape and the
  Swampland},'' \href{http://dx.doi.org/10.1016/j.nuclphysb.2006.10.033}{{\em
  Nucl. Phys. B} {\bf 766} (2007)  21--33},
  \href{http://arxiv.org/abs/hep-th/0605264}{{\tt arXiv:hep-th/0605264}}.

\bibitem{Garg:2018reu}
S.~K. Garg and C.~Krishnan, ``{Bounds on Slow Roll and the de Sitter
  Swampland},'' \href{http://dx.doi.org/10.1007/JHEP11(2019)075}{{\em JHEP}
  {\bf 11} (2019)  075}, \href{http://arxiv.org/abs/1807.05193}{{\tt
  arXiv:1807.05193 [hep-th]}}.

\bibitem{Ooguri:2018wrx}
H.~Ooguri, E.~Palti, G.~Shiu, and C.~Vafa, ``{Distance and de Sitter
  Conjectures on the Swampland},''
  \href{http://dx.doi.org/10.1016/j.physletb.2018.11.018}{{\em Phys. Lett. B}
  {\bf 788} (2019)  180--184}, \href{http://arxiv.org/abs/1810.05506}{{\tt
  arXiv:1810.05506 [hep-th]}}.

\bibitem{Agrawal:2018own}
P.~Agrawal, G.~Obied, P.~J. Steinhardt, and C.~Vafa, ``{On the Cosmological
  Implications of the String Swampland},''
  \href{http://dx.doi.org/10.1016/j.physletb.2018.07.040}{{\em Phys. Lett. B}
  {\bf 784} (2018)  271--276}, \href{http://arxiv.org/abs/1806.09718}{{\tt
  arXiv:1806.09718 [hep-th]}}.

\bibitem{Vilenkin:1983xq}
A.~Vilenkin, ``{The Birth of Inflationary Universes},''
  \href{http://dx.doi.org/10.1103/PhysRevD.27.2848}{{\em Phys. Rev. D} {\bf 27}
  (1983)  2848}.

\bibitem{Winitzki:2008zz}
S.~Winitzki, \href{http://dx.doi.org/10.1142/6923}{{\em {Eternal inflation}}}.
\newblock 2008.

\bibitem{Guth:2007ng}
A.~H. Guth, ``{Eternal inflation and its implications},''
  \href{http://dx.doi.org/10.1088/1751-8113/40/25/S25}{{\em J. Phys. A} {\bf
  40} (2007)  6811--6826}, \href{http://arxiv.org/abs/hep-th/0702178}{{\tt
  arXiv:hep-th/0702178}}.

\bibitem{Linde:1986fc}
A.~D. Linde, ``{ETERNAL CHAOTIC INFLATION},''
  \href{http://dx.doi.org/10.1142/S0217732386000129}{{\em Mod. Phys. Lett. A}
  {\bf 1} (1986)  81}.

\bibitem{Linde:1986fd}
A.~D. Linde, ``{Eternally Existing Selfreproducing Chaotic Inflationary
  Universe},'' \href{http://dx.doi.org/10.1016/0370-2693(86)90611-8}{{\em Phys.
  Lett. B} {\bf 175} (1986)  395--400}.

\bibitem{Weinberg:1987dv}
S.~Weinberg, ``{Anthropic Bound on the Cosmological Constant},''
\href{http://dx.doi.org/10.1103/PhysRevLett.59.2607}{{\em Phys. Rev. Lett.}
  {\bf 59} (1987)  2607}.
%%CITATION = PRLTA,59,2607;%%.

\bibitem{Agrawal:1997gf}
V.~Agrawal, S.~M. Barr, J.~F. Donoghue, and D.~Seckel, ``{Viable range of the
  mass scale of the standard model},''
  \href{http://dx.doi.org/10.1103/PhysRevD.57.5480}{{\em Phys. Rev.} {\bf D57}
  (1998)  5480--5492},
\href{http://arxiv.org/abs/hep-ph/9707380}{{\tt arXiv:hep-ph/9707380
  [hep-ph]}}.
%%CITATION = HEP-PH/9707380;%%.

\bibitem{Arkani-Hamed:2004ymt}
N.~Arkani-Hamed and S.~Dimopoulos, ``{Supersymmetric unification without low
  energy supersymmetry and signatures for fine-tuning at the LHC},''
  \href{http://dx.doi.org/10.1088/1126-6708/2005/06/073}{{\em JHEP} {\bf 06}
  (2005)  073}, \href{http://arxiv.org/abs/hep-th/0405159}{{\tt
  arXiv:hep-th/0405159}}.

\bibitem{Geller:2018xvz}
M.~Geller, Y.~Hochberg, and E.~Kuflik, ``{Inflating to the Weak Scale},''
  \href{http://dx.doi.org/10.1103/PhysRevLett.122.191802}{{\em Phys. Rev.
  Lett.} {\bf 122} (2019) no.~19, 191802},
\href{http://arxiv.org/abs/1809.07338}{{\tt arXiv:1809.07338 [hep-ph]}}.
%%CITATION = ARXIV:1809.07338;%%.

\bibitem{Giudice:2019iwl}
G.~F. Giudice, A.~Kehagias, and A.~Riotto, ``{The Selfish Higgs},''
\href{http://arxiv.org/abs/1907.05370}{{\tt arXiv:1907.05370 [hep-ph]}}.
%%CITATION = ARXIV:1907.05370;%%.

\bibitem{Arkani-Hamed:2020yna}
N.~Arkani-Hamed, R.~T. D'Agnolo, and H.~D. Kim, ``{The Weak Scale as a
  Trigger},'' \href{http://arxiv.org/abs/2012.04652}{{\tt arXiv:2012.04652
  [hep-ph]}}.

\bibitem{Strumia:2020bdy}
A.~Strumia and D.~Teresi, ``{Relaxing the Higgs mass and its vacuum energy by
  living at the top of the potential},''
  \href{http://dx.doi.org/10.1103/PhysRevD.101.115002}{{\em Phys. Rev. D} {\bf
  101} (2020) no.~11, 115002}, \href{http://arxiv.org/abs/2002.02463}{{\tt
  arXiv:2002.02463 [hep-ph]}}.

\bibitem{Csaki:2020zqz}
C.~Cs\'aki, R.~T. D'Agnolo, M.~Geller, and A.~Ismail, ``{Crunching Dilaton,
  Hidden Naturalness},''
  \href{http://dx.doi.org/10.1103/PhysRevLett.126.091801}{{\em Phys. Rev.
  Lett.} {\bf 126} (2021)  091801}, \href{http://arxiv.org/abs/2007.14396}{{\tt
  arXiv:2007.14396 [hep-ph]}}.

\bibitem{TitoDAgnolo:2021nhd}
R.~T. D'Agnolo and D.~Teresi, ``{Sliding Naturalness},''
  \href{http://arxiv.org/abs/2106.04591}{{\tt arXiv:2106.04591 [hep-ph]}}.

\bibitem{TitoDAgnolo:2021pjo}
R.~Tito~D'Agnolo and D.~Teresi, ``{Sliding Naturalness: Cosmological Selection
  of the Weak Scale},'' \href{http://arxiv.org/abs/2109.13249}{{\tt
  arXiv:2109.13249 [hep-ph]}}.

\bibitem{Giudice:2021viw}
G.~F. Giudice, M.~McCullough, and T.~You, ``{Self-Organised Localisation},''
  \href{http://arxiv.org/abs/2105.08617}{{\tt arXiv:2105.08617 [hep-ph]}}.

\bibitem{Khoury:2021zao}
J.~Khoury and T.~Steingasser, ``{Gauge hierarchy from electroweak vacuum
  metastability},'' \href{http://dx.doi.org/10.1103/PhysRevD.105.055031}{{\em
  Phys. Rev. D} {\bf 105} (2022) no.~5, 055031},
  \href{http://arxiv.org/abs/2108.09315}{{\tt arXiv:2108.09315 [hep-ph]}}.

\bibitem{Martin:1997ns}
S.~P. Martin, ``{A Supersymmetry primer},''
  \href{http://dx.doi.org/10.1142/9789812839657_0001}{{\em Adv. Ser. Direct.
  High Energy Phys.} {\bf 18} (1998)  1--98},
  \href{http://arxiv.org/abs/hep-ph/9709356}{{\tt arXiv:hep-ph/9709356}}.

\bibitem{Panico:2015jxa}
G.~Panico and A.~Wulzer,
  \href{http://dx.doi.org/10.1007/978-3-319-22617-0}{{\em {The Composite
  Nambu-Goldstone Higgs}}}, vol.~913.
\newblock Springer, 2016.
\newblock \href{http://arxiv.org/abs/1506.01961}{{\tt arXiv:1506.01961
  [hep-ph]}}.

\bibitem{Linde:1994gy}
A.~D. Linde, D.~A. Linde, and A.~Mezhlumian, ``{Do we live in the center of the
  world?},'' \href{http://dx.doi.org/10.1016/0370-2693(94)01641-O}{{\em Phys.
  Lett. B} {\bf 345} (1995)  203--210},
  \href{http://arxiv.org/abs/hep-th/9411111}{{\tt arXiv:hep-th/9411111}}.

\bibitem{Vilenkin:1998kr}
A.~Vilenkin, ``{Unambiguous probabilities in an eternally inflating
  universe},'' \href{http://dx.doi.org/10.1103/PhysRevLett.81.5501}{{\em Phys.
  Rev. Lett.} {\bf 81} (1998)  5501--5504},
  \href{http://arxiv.org/abs/hep-th/9806185}{{\tt arXiv:hep-th/9806185}}.

\bibitem{Peebles:1987ek}
P.~J.~E. Peebles and B.~Ratra, ``{Cosmology with a Time Variable Cosmological
  Constant},'' \href{http://dx.doi.org/10.1086/185100}{{\em Astrophys. J.
  Lett.} {\bf 325} (1988)  L17}.

\bibitem{Ratra:1987rm}
B.~Ratra and P.~J.~E. Peebles, ``{Cosmological Consequences of a Rolling
  Homogeneous Scalar Field},''
  \href{http://dx.doi.org/10.1103/PhysRevD.37.3406}{{\em Phys. Rev. D} {\bf 37}
  (1988)  3406}.

\bibitem{Frieman:1995pm}
J.~A. Frieman, C.~T. Hill, A.~Stebbins, and I.~Waga, ``{Cosmology with
  ultralight pseudo Nambu-Goldstone bosons},''
  \href{http://dx.doi.org/10.1103/PhysRevLett.75.2077}{{\em Phys. Rev. Lett.}
  {\bf 75} (1995)  2077--2080},
  \href{http://arxiv.org/abs/astro-ph/9505060}{{\tt arXiv:astro-ph/9505060}}.

\bibitem{Ferreira:1997au}
P.~G. Ferreira and M.~Joyce, ``{Structure formation with a selftuning scalar
  field},'' \href{http://dx.doi.org/10.1103/PhysRevLett.79.4740}{{\em Phys.
  Rev. Lett.} {\bf 79} (1997)  4740--4743},
  \href{http://arxiv.org/abs/astro-ph/9707286}{{\tt arXiv:astro-ph/9707286}}.

\bibitem{Caldwell:1997ii}
R.~R. Caldwell, R.~Dave, and P.~J. Steinhardt, ``{Cosmological imprint of an
  energy component with general equation of state},''
  \href{http://dx.doi.org/10.1103/PhysRevLett.80.1582}{{\em Phys. Rev. Lett.}
  {\bf 80} (1998)  1582--1585},
  \href{http://arxiv.org/abs/astro-ph/9708069}{{\tt arXiv:astro-ph/9708069}}.

\bibitem{DESI:2024mwx}
{\bf DESI} Collaboration, A.~G. Adame {\em et al.}, ``{DESI 2024 VI:
  Cosmological Constraints from the Measurements of Baryon Acoustic
  Oscillations},'' \href{http://arxiv.org/abs/2404.03002}{{\tt arXiv:2404.03002
  [astro-ph.CO]}}.

\bibitem{Dvali:2013eja}
G.~Dvali and C.~Gomez, ``{Quantum Compositeness of Gravity: Black Holes, AdS
  and Inflation},'' \href{http://dx.doi.org/10.1088/1475-7516/2014/01/023}{{\em
  JCAP} {\bf 01} (2014)  023}, \href{http://arxiv.org/abs/1312.4795}{{\tt
  arXiv:1312.4795 [hep-th]}}.

\bibitem{Dvali:2014gua}
G.~Dvali and C.~Gomez, ``{Quantum Exclusion of Positive Cosmological
  Constant?},'' \href{http://dx.doi.org/10.1002/andp.201500216}{{\em Annalen
  Phys.} {\bf 528} (2016)  68--73}, \href{http://arxiv.org/abs/1412.8077}{{\tt
  arXiv:1412.8077 [hep-th]}}.

\bibitem{Dvali:2017eba}
G.~Dvali, C.~Gomez, and S.~Zell, ``{Quantum Break-Time of de Sitter},''
  \href{http://dx.doi.org/10.1088/1475-7516/2017/06/028}{{\em JCAP} {\bf 06}
  (2017)  028}, \href{http://arxiv.org/abs/1701.08776}{{\tt arXiv:1701.08776
  [hep-th]}}.

\bibitem{Dvali:2020etd}
G.~Dvali, ``{$S$-Matrix and Anomaly of de Sitter},''
  \href{http://dx.doi.org/10.3390/sym13010003}{{\em Symmetry} {\bf 13} (2020)
  no.~1, 3}, \href{http://arxiv.org/abs/2012.02133}{{\tt arXiv:2012.02133
  [hep-th]}}.

\bibitem{Dvali:2021kxt}
G.~Dvali, ``{On $S$-Matrix Exclusion of de Sitter and Naturalness},''
  \href{http://arxiv.org/abs/2105.08411}{{\tt arXiv:2105.08411 [hep-th]}}.

\bibitem{Wang:2019eym}
Z.~Wang, R.~Brandenberger, and L.~Heisenberg, ``{Eternal Inflation, Entropy
  Bounds and the Swampland},''
  \href{http://dx.doi.org/10.1140/epjc/s10052-020-8412-x}{{\em Eur. Phys. J. C}
  {\bf 80} (2020) no.~9, 864}, \href{http://arxiv.org/abs/1907.08943}{{\tt
  arXiv:1907.08943 [hep-th]}}.

\bibitem{Kleban:2012ph}
M.~Kleban and M.~Schillo, ``{Spatial Curvature Falsifies Eternal Inflation},''
  \href{http://dx.doi.org/10.1088/1475-7516/2012/06/029}{{\em JCAP} {\bf 06}
  (2012)  029}, \href{http://arxiv.org/abs/1202.5037}{{\tt arXiv:1202.5037
  [astro-ph.CO]}}.

\bibitem{Zeldovic_Grisuchuk}
L.~P. {Grishchuk} and I.~B. {Zeldovich}, ``{Long-wavelength perturbations of a
  Friedmann universe, and anisotropy of the microwave background
  radiation},''{\em sovast} {\bf 22} (Apr., 1978)  125--129.

\bibitem{Kinney:2014jya}
W.~H. Kinney and K.~Freese, ``{Negative running can prevent eternal
  inflation},'' \href{http://dx.doi.org/10.1088/1475-7516/2015/01/040}{{\em
  JCAP} {\bf 01} (2015)  040}, \href{http://arxiv.org/abs/1404.4614}{{\tt
  arXiv:1404.4614 [astro-ph.CO]}}.

\bibitem{Montefalcone:2023izs}
G.~Montefalcone, R.~O.~Ramos, G.~S.~Vicente, and K.~Freese, ``{Defying eternal
  inflation in warm inflation with a negative running},''
  \href{http://dx.doi.org/10.1088/1475-7516/2024/02/006}{{\em JCAP} {\bf 02}
  (2024)  006}, \href{http://arxiv.org/abs/2311.03487}{{\tt arXiv:2311.03487
  [astro-ph.CO]}}.

\bibitem{Planck:2018jri}
{\bf Planck} Collaboration, Y.~Akrami {\em et al.}, ``{Planck 2018 results. X.
  Constraints on inflation},''
  \href{http://dx.doi.org/10.1051/0004-6361/201833887}{{\em Astron. Astrophys.}
  {\bf 641} (2020)  A10}, \href{http://arxiv.org/abs/1807.06211}{{\tt
  arXiv:1807.06211 [astro-ph.CO]}}.

\bibitem{Dimopoulos:2018upl}
K.~Dimopoulos, ``{Steep Eternal Inflation and the Swampland},''
  \href{http://dx.doi.org/10.1103/PhysRevD.98.123516}{{\em Phys. Rev. D} {\bf
  98} (2018) no.~12, 123516}, \href{http://arxiv.org/abs/1810.03438}{{\tt
  arXiv:1810.03438 [gr-qc]}}.

\bibitem{Matsui:2018bsy}
H.~Matsui and F.~Takahashi, ``{Eternal Inflation and Swampland Conjectures},''
  \href{http://dx.doi.org/10.1103/PhysRevD.99.023533}{{\em Phys. Rev. D} {\bf
  99} (2019) no.~2, 023533}, \href{http://arxiv.org/abs/1807.11938}{{\tt
  arXiv:1807.11938 [hep-th]}}.

\bibitem{Blanco-Pillado:2019tdf}
J.~J. Blanco-Pillado, H.~Deng, and A.~Vilenkin, ``{Eternal Inflation in Swampy
  Landscapes},'' \href{http://dx.doi.org/10.1088/1475-7516/2020/05/014}{{\em
  JCAP} {\bf 05} (2020)  014}, \href{http://arxiv.org/abs/1909.00068}{{\tt
  arXiv:1909.00068 [gr-qc]}}.

\bibitem{Kinney:2018kew}
W.~H. Kinney, ``{Eternal Inflation and the Refined Swampland Conjecture},''
  \href{http://dx.doi.org/10.1103/PhysRevLett.122.081302}{{\em Phys. Rev.
  Lett.} {\bf 122} (2019) no.~8, 081302},
  \href{http://arxiv.org/abs/1811.11698}{{\tt arXiv:1811.11698 [astro-ph.CO]}}.

\bibitem{Brahma:2019iyy}
S.~Brahma and S.~Shandera, ``{Stochastic eternal inflation is in the
  swampland},'' \href{http://dx.doi.org/10.1007/JHEP11(2019)016}{{\em JHEP}
  {\bf 11} (2019)  016}, \href{http://arxiv.org/abs/1904.10979}{{\tt
  arXiv:1904.10979 [hep-th]}}.

\bibitem{Lin:2019fdk}
C.-M. Lin, ``{Topological Eternal Hilltop Inflation and the Swampland
  Criteria},'' \href{http://dx.doi.org/10.1088/1475-7516/2020/06/015}{{\em
  JCAP} {\bf 06} (2020)  015}, \href{http://arxiv.org/abs/1912.00749}{{\tt
  arXiv:1912.00749 [hep-th]}}.

\bibitem{Martin:2000xs}
J.~Martin and R.~H. Brandenberger, ``{The TransPlanckian problem of
  inflationary cosmology},''
  \href{http://dx.doi.org/10.1103/PhysRevD.63.123501}{{\em Phys. Rev. D} {\bf
  63} (2001)  123501}, \href{http://arxiv.org/abs/hep-th/0005209}{{\tt
  arXiv:hep-th/0005209}}.

\bibitem{Bedroya:2019tba}
A.~Bedroya, R.~Brandenberger, M.~Loverde, and C.~Vafa, ``{Trans-Planckian
  Censorship and Inflationary Cosmology},''
  \href{http://dx.doi.org/10.1103/PhysRevD.101.103502}{{\em Phys. Rev. D} {\bf
  101} (2020) no.~10, 103502}, \href{http://arxiv.org/abs/1909.11106}{{\tt
  arXiv:1909.11106 [hep-th]}}.

\bibitem{Bedroya:2019snp}
A.~Bedroya and C.~Vafa, ``{Trans-Planckian Censorship and the Swampland},''
  \href{http://dx.doi.org/10.1007/JHEP09(2020)123}{{\em JHEP} {\bf 09} (2020)
  123}, \href{http://arxiv.org/abs/1909.11063}{{\tt arXiv:1909.11063
  [hep-th]}}.

\bibitem{Burgess:2020nec}
C.~P. Burgess, S.~P. de~Alwis, and F.~Quevedo, ``{Cosmological Trans-Planckian
  Conjectures are not Effective},''
  \href{http://dx.doi.org/10.1088/1475-7516/2021/05/037}{{\em JCAP} {\bf 05}
  (2021)  037}, \href{http://arxiv.org/abs/2011.03069}{{\tt arXiv:2011.03069
  [hep-th]}}.

\bibitem{Komissarov:2022gax}
I.~Komissarov, A.~Nicolis, and J.~Staunton, ``{Cosmology as a weak
  gravitational field and the trans-Planckian problem},''
  \href{http://dx.doi.org/10.1007/JHEP05(2023)216}{{\em JHEP} {\bf 05} (2023)
  216}, \href{http://arxiv.org/abs/2210.11508}{{\tt arXiv:2210.11508
  [hep-th]}}.

\bibitem{Hawking:1981fz}
S.~W. Hawking and I.~G. Moss, ``{Supercooled Phase Transitions in the Very
  Early Universe},'' \href{http://dx.doi.org/10.1016/0370-2693(82)90946-7}{{\em
  Phys. Lett. B} {\bf 110} (1982)  35--38}.

\bibitem{Blanco-Pillado:2019xny}
J.~J. Blanco-Pillado, H.~Deng, and A.~Vilenkin, ``{Flyover vacuum decay},''
  \href{http://dx.doi.org/10.1088/1475-7516/2019/12/001}{{\em JCAP} {\bf 12}
  (2019)  001}, \href{http://arxiv.org/abs/1906.09657}{{\tt arXiv:1906.09657
  [hep-th]}}.

\bibitem{Lee:1987qc}
K.-M. Lee and E.~J. Weinberg, ``{Decay of the True Vacuum in Curved
  Space-time},'' \href{http://dx.doi.org/10.1103/PhysRevD.36.1088}{{\em Phys.
  Rev. D} {\bf 36} (1987)  1088}.

\bibitem{PhysRevD.36.2919}
V.~A. Berezin, V.~A. Kuzmin, and I.~I. Tkachev,
  \href{http://dx.doi.org/10.1103/PhysRevD.36.2919}{``Dynamics of bubbles in
  general relativity,''{\em Phys. Rev. D} {\bf 36} (Nov, 1987)  2919--2944}.
  \url{https://link.aps.org/doi/10.1103/PhysRevD.36.2919}.

\bibitem{BEREZIN198323}
V.~Berezin, V.~Kuzmin, and I.~Tkachev, ``New vacuum formation in the
  universe,''
  \href{http://dx.doi.org/https://doi.org/10.1016/0370-2693(83)91055-9}{{\em
  Physics Letters B} {\bf 130} (1983) no.~1, 23--27}.
  \url{https://www.sciencedirect.com/science/article/pii/0370269383910559}.

\bibitem{Berezin:112638}
World Scientific, {\em {}}.
\newblock World Scientific, Singapore, 1988.
\newblock \url{https://cds.cern.ch/record/112638}.

\bibitem{Basu:1991ig}
R.~Basu, A.~H. Guth, and A.~Vilenkin, ``{Quantum creation of topological
  defects during inflation},''
  \href{http://dx.doi.org/10.1103/PhysRevD.44.340}{{\em Phys. Rev. D} {\bf 44}
  (1991)  340--351}.

\bibitem{Sato:1981gv}
K.~Sato, H.~Kodama, M.~Sasaki, and K.-i. Maeda, ``{Multiproduction of Universes
  by First Order Phase Transition of a Vacuum},''
  \href{http://dx.doi.org/10.1016/0370-2693(82)91152-2}{{\em Phys. Lett. B}
  {\bf 108} (1982)  103--107}.

\bibitem{Blau:1986cw}
S.~K. Blau, E.~I. Guendelman, and A.~H. Guth, ``{The Dynamics of False Vacuum
  Bubbles},'' \href{http://dx.doi.org/10.1103/PhysRevD.35.1747}{{\em Phys. Rev.
  D} {\bf 35} (1987)  1747}.

\bibitem{Ellis:2013dla}
G.~F.~R. Ellis and R.~Goswami, ``{Variations on Birkhoff`s theorem},''
  \href{http://dx.doi.org/10.1007/s10714-013-1568-z}{{\em Gen. Rel. Grav.} {\bf
  45} (2013)  2123--2142}, \href{http://arxiv.org/abs/1304.3253}{{\tt
  arXiv:1304.3253 [gr-qc]}}.

\bibitem{Israel:1966rt}
W.~Israel, ``{Singular hypersurfaces and thin shells in general relativity},''
  \href{http://dx.doi.org/10.1007/BF02710419}{{\em Nuovo Cim. B} {\bf 44S10}
  (1966)  1}. [Erratum: Nuovo Cim.B 48, 463 (1967)].

\bibitem{Coleman:1977py}
S.~R. Coleman, ``{The Fate of the False Vacuum. 1. Semiclassical Theory},''
  \href{http://dx.doi.org/10.1103/PhysRevD.16.1248}{{\em Phys. Rev. D} {\bf 15}
  (1977)  2929--2936}. [Erratum: Phys.Rev.D 16, 1248 (1977)].

\bibitem{Coleman:1980aw}
S.~R. Coleman and F.~De~Luccia, ``{Gravitational Effects on and of Vacuum
  Decay},'' \href{http://dx.doi.org/10.1103/PhysRevD.21.3305}{{\em Phys. Rev.
  D} {\bf 21} (1980)  3305}.

\bibitem{Farhi:1989yr}
E.~Farhi, A.~H. Guth, and J.~Guven, ``{Is It Possible to Create a Universe in
  the Laboratory by Quantum Tunneling?},''
  \href{http://dx.doi.org/10.1016/0550-3213(90)90357-J}{{\em Nucl. Phys. B}
  {\bf 339} (1990)  417--490}.

\bibitem{Fischler:1989se}
W.~Fischler, D.~Morgan, and J.~Polchinski, ``{Quantum Nucleation of False
  Vacuum Bubbles},'' \href{http://dx.doi.org/10.1103/PhysRevD.41.2638}{{\em
  Phys. Rev. D} {\bf 41} (1990)  2638}.

\bibitem{Farhi:1986ty}
E.~Farhi and A.~H. Guth, ``{An Obstacle to Creating a Universe in the
  Laboratory},'' \href{http://dx.doi.org/10.1016/0370-2693(87)90429-1}{{\em
  Phys. Lett. B} {\bf 183} (1987)  149--155}.

\bibitem{PhysRevD.45.3469}
J.~Garriga and A.~Vilenkin,
  \href{http://dx.doi.org/10.1103/PhysRevD.45.3469}{``Quantum fluctuations on
  domain walls, strings, and vacuum bubbles,''{\em Phys. Rev. D} {\bf 45} (May,
  1992)  3469--3486}. \url{https://link.aps.org/doi/10.1103/PhysRevD.45.3469}.

\bibitem{Deng:2017uwc}
H.~Deng and A.~Vilenkin, ``{Primordial black hole formation by vacuum
  bubbles},'' \href{http://dx.doi.org/10.1088/1475-7516/2017/12/044}{{\em JCAP}
  {\bf 12} (2017)  044}, \href{http://arxiv.org/abs/1710.02865}{{\tt
  arXiv:1710.02865 [gr-qc]}}.

\bibitem{Bloch:2019bvc}
I.~M. Bloch, C.~Cs\'aki, M.~Geller, and T.~Volansky, ``{Crunching away the
  cosmological constant problem: dynamical selection of a small $\Lambda$},''
  \href{http://dx.doi.org/10.1007/JHEP12(2020)191}{{\em JHEP} {\bf 12} (2020)
  191}, \href{http://arxiv.org/abs/1912.08840}{{\tt arXiv:1912.08840
  [hep-ph]}}.

\bibitem{Aragam:2019omo}
V.~Aragam, S.~Paban, and R.~Rosati, ``{Multi-field Inflation in High-Slope
  Potentials},'' \href{http://dx.doi.org/10.1088/1475-7516/2020/04/022}{{\em
  JCAP} {\bf 04} (2020)  022}, \href{http://arxiv.org/abs/1905.07495}{{\tt
  arXiv:1905.07495 [hep-th]}}.

\bibitem{Kehagias:2018uem}
A.~Kehagias and A.~Riotto, ``{A note on Inflation and the Swampland},''
  \href{http://dx.doi.org/10.1002/prop.201800052}{{\em Fortsch. Phys.} {\bf 66}
  (2018) no.~10, 1800052}, \href{http://arxiv.org/abs/1807.05445}{{\tt
  arXiv:1807.05445 [hep-th]}}.

\bibitem{Bravo:2020wdr}
R.~Bravo, G.~A. Palma, and S.~Riquelme, ``{A Tip for Landscape Riders:
  Multi-Field Inflation Can Fulfill the Swampland Distance Conjecture},''
  \href{http://dx.doi.org/10.1088/1475-7516/2020/02/004}{{\em JCAP} {\bf 02}
  (2020)  004}, \href{http://arxiv.org/abs/1906.05772}{{\tt arXiv:1906.05772
  [hep-th]}}.

\bibitem{Aragam:2021scu}
V.~Aragam, R.~Chiovoloni, S.~Paban, R.~Rosati, and I.~Zavala, ``{Rapid-turn
  inflation in supergravity is rare and tachyonic},''
  \href{http://dx.doi.org/10.1088/1475-7516/2022/03/002}{{\em JCAP} {\bf 03}
  (2022) no.~03, 002}, \href{http://arxiv.org/abs/2110.05516}{{\tt
  arXiv:2110.05516 [hep-th]}}.

\bibitem{Achucarro:2018vey}
A.~Ach\'ucarro and G.~A. Palma, ``{The string swampland constraints require
  multi-field inflation},''
  \href{http://dx.doi.org/10.1088/1475-7516/2019/02/041}{{\em JCAP} {\bf 02}
  (2019)  041}, \href{http://arxiv.org/abs/1807.04390}{{\tt arXiv:1807.04390
  [hep-th]}}.

\bibitem{Linde:1993cn}
A.~D. Linde, ``{Hybrid inflation},''
  \href{http://dx.doi.org/10.1103/PhysRevD.49.748}{{\em Phys. Rev. D} {\bf 49}
  (1994)  748--754}, \href{http://arxiv.org/abs/astro-ph/9307002}{{\tt
  arXiv:astro-ph/9307002}}.

\bibitem{Kallosh:2013yoa}
R.~Kallosh, A.~Linde, and D.~Roest, ``{Superconformal Inflationary
  $\alpha$-Attractors},'' \href{http://dx.doi.org/10.1007/JHEP11(2013)198}{{\em
  JHEP} {\bf 11} (2013)  198}, \href{http://arxiv.org/abs/1311.0472}{{\tt
  arXiv:1311.0472 [hep-th]}}.

\bibitem{Roest:2015qya}
D.~Roest and M.~Scalisi, ``{Cosmological attractors from
  \ensuremath{\alpha}-scale supergravity},''
  \href{http://dx.doi.org/10.1103/PhysRevD.92.043525}{{\em Phys. Rev. D} {\bf
  92} (2015)  043525}, \href{http://arxiv.org/abs/1503.07909}{{\tt
  arXiv:1503.07909 [hep-th]}}.

\bibitem{Galante:2014ifa}
M.~Galante, R.~Kallosh, A.~Linde, and D.~Roest, ``{Unity of Cosmological
  Inflation Attractors},''
  \href{http://dx.doi.org/10.1103/PhysRevLett.114.141302}{{\em Phys. Rev.
  Lett.} {\bf 114} (2015) no.~14, 141302},
  \href{http://arxiv.org/abs/1412.3797}{{\tt arXiv:1412.3797 [hep-th]}}.

\bibitem{Scalisi:2018eaz}
M.~Scalisi and I.~Valenzuela, ``{Swampland distance conjecture, inflation and
  $\alpha$-attractors},'' \href{http://dx.doi.org/10.1007/JHEP08(2019)160}{{\em
  JHEP} {\bf 08} (2019)  160}, \href{http://arxiv.org/abs/1812.07558}{{\tt
  arXiv:1812.07558 [hep-th]}}.

\bibitem{Berghaus:2019whh}
K.~V. Berghaus, P.~W. Graham, and D.~E. Kaplan, ``{Minimal Warm Inflation},''
  \href{http://dx.doi.org/10.1088/1475-7516/2020/03/034}{{\em JCAP} {\bf 03}
  (2020)  034}, \href{http://arxiv.org/abs/1910.07525}{{\tt arXiv:1910.07525
  [hep-ph]}}. [Erratum: JCAP 10, E02 (2023)].

\bibitem{Moore:2010jd}
G.~D. Moore and M.~Tassler, ``{The Sphaleron Rate in SU(N) Gauge Theory},''
  \href{http://dx.doi.org/10.1007/JHEP02(2011)105}{{\em JHEP} {\bf 02} (2011)
  105}, \href{http://arxiv.org/abs/1011.1167}{{\tt arXiv:1011.1167 [hep-ph]}}.

\bibitem{Laine:2016hma}
M.~Laine and A.~Vuorinen,
  \href{http://dx.doi.org/10.1007/978-3-319-31933-9}{{\em {Basics of Thermal
  Field Theory}}}, vol.~925.
\newblock Springer, 2016.
\newblock \href{http://arxiv.org/abs/1701.01554}{{\tt arXiv:1701.01554
  [hep-ph]}}.

\bibitem{Graham:2009bf}
C.~Graham and I.~G. Moss, ``{Density fluctuations from warm inflation},''
  \href{http://dx.doi.org/10.1088/1475-7516/2009/07/013}{{\em JCAP} {\bf 07}
  (2009)  013}, \href{http://arxiv.org/abs/0905.3500}{{\tt arXiv:0905.3500
  [astro-ph.CO]}}.

\bibitem{Mirbabayi:2022cbt}
M.~Mirbabayi and A.~Gruzinov, ``{Shapes of non-Gaussianity in warm
  inflation},'' \href{http://dx.doi.org/10.1088/1475-7516/2023/02/012}{{\em
  JCAP} {\bf 02} (2023)  012}, \href{http://arxiv.org/abs/2205.13227}{{\tt
  arXiv:2205.13227 [astro-ph.CO]}}.

\bibitem{Berghaus:2024zfg}
K.~V. Berghaus, M.~Forslund, and M.~V. Guevarra, ``{Minimal warm inflation with
  a heavy QCD axion},'' \href{http://arxiv.org/abs/2402.13535}{{\tt
  arXiv:2402.13535 [hep-ph]}}.

\bibitem{McAllister:2016vzi}
L.~McAllister, P.~Schwaller, G.~Servant, J.~Stout, and A.~Westphal, ``{Runaway
  Relaxion Monodromy},'' \href{http://dx.doi.org/10.1007/JHEP02(2018)124}{{\em
  JHEP} {\bf 02} (2018)  124}, \href{http://arxiv.org/abs/1610.05320}{{\tt
  arXiv:1610.05320 [hep-th]}}.

\bibitem{Graham:2015cka}
P.~W. Graham, D.~E. Kaplan, and S.~Rajendran, ``{Cosmological Relaxation of the
  Electroweak Scale},''
  \href{http://dx.doi.org/10.1103/PhysRevLett.115.221801}{{\em Phys. Rev.
  Lett.} {\bf 115} (2015) no.~22, 221801},
\href{http://arxiv.org/abs/1504.07551}{{\tt arXiv:1504.07551 [hep-ph]}}.
%%CITATION = ARXIV:1504.07551;%%.

\bibitem{Yokoyama:1998ju}
J.~Yokoyama and A.~D. Linde, ``{Is warm inflation possible?},''
  \href{http://dx.doi.org/10.1103/PhysRevD.60.083509}{{\em Phys. Rev. D} {\bf
  60} (1999)  083509}, \href{http://arxiv.org/abs/hep-ph/9809409}{{\tt
  arXiv:hep-ph/9809409}}.

\bibitem{Berera:1998px}
A.~Berera, M.~Gleiser, and R.~O. Ramos, ``{A First principles warm inflation
  model that solves the cosmological horizon / flatness problems},''
  \href{http://dx.doi.org/10.1103/PhysRevLett.83.264}{{\em Phys. Rev. Lett.}
  {\bf 83} (1999)  264--267}, \href{http://arxiv.org/abs/hep-ph/9809583}{{\tt
  arXiv:hep-ph/9809583}}.

\bibitem{Bastero-Gil:2016qru}
M.~Bastero-Gil, A.~Berera, R.~O. Ramos, and J.~G. Rosa, ``{Warm Little
  Inflaton},'' \href{http://dx.doi.org/10.1103/PhysRevLett.117.151301}{{\em
  Phys. Rev. Lett.} {\bf 117} (2016) no.~15, 151301},
  \href{http://arxiv.org/abs/1604.08838}{{\tt arXiv:1604.08838 [hep-ph]}}.

\bibitem{Bastero-Gil:2019gao}
M.~Bastero-Gil, A.~Berera, R.~O. Ramos, and J.~a.~G. Rosa, ``{Towards a
  reliable effective field theory of inflation},''
  \href{http://dx.doi.org/10.1016/j.physletb.2020.136055}{{\em Phys. Lett. B}
  {\bf 813} (2021)  136055}, \href{http://arxiv.org/abs/1907.13410}{{\tt
  arXiv:1907.13410 [hep-ph]}}.

\bibitem{Motaharfar:2018zyb}
M.~Motaharfar, V.~Kamali, and R.~O. Ramos, ``{Warm inflation as a way out of
  the swampland},'' \href{http://dx.doi.org/10.1103/PhysRevD.99.063513}{{\em
  Phys. Rev. D} {\bf 99} (2019) no.~6, 063513},
  \href{http://arxiv.org/abs/1810.02816}{{\tt arXiv:1810.02816 [astro-ph.CO]}}.

\bibitem{Das:2018rpg}
S.~Das, ``{Warm Inflation in the light of Swampland Criteria},''
  \href{http://dx.doi.org/10.1103/PhysRevD.99.063514}{{\em Phys. Rev. D} {\bf
  99} (2019) no.~6, 063514}, \href{http://arxiv.org/abs/1810.05038}{{\tt
  arXiv:1810.05038 [hep-th]}}.

\bibitem{Brandenberger:2020oav}
R.~Brandenberger, V.~Kamali, and R.~O. Ramos, ``{Strengthening the de Sitter
  swampland conjecture in warm inflation},''
  \href{http://dx.doi.org/10.1007/JHEP08(2020)127}{{\em JHEP} {\bf 08} (2020)
  127}, \href{http://arxiv.org/abs/2002.04925}{{\tt arXiv:2002.04925
  [hep-th]}}.

\bibitem{Das:2020xmh}
S.~Das and R.~O. Ramos, ``{Runaway potentials in warm inflation satisfying the
  swampland conjectures},''
  \href{http://dx.doi.org/10.1103/PhysRevD.102.103522}{{\em Phys. Rev. D} {\bf
  102} (2020) no.~10, 103522}, \href{http://arxiv.org/abs/2007.15268}{{\tt
  arXiv:2007.15268 [hep-th]}}.

\bibitem{Arkani-Hamed:2005zuc}
N.~Arkani-Hamed, S.~Dimopoulos, and S.~Kachru, ``{Predictive landscapes and new
  physics at a TeV},'' \href{http://arxiv.org/abs/hep-th/0501082}{{\tt
  arXiv:hep-th/0501082}}.

\bibitem{Callan:1977pt}
C.~G. Callan, Jr. and S.~R. Coleman, ``{The Fate of the False Vacuum. 2. First
  Quantum Corrections},''
  \href{http://dx.doi.org/10.1103/PhysRevD.16.1762}{{\em Phys. Rev. D} {\bf 16}
  (1977)  1762--1768}.

\bibitem{Bousso:2000xa}
R.~Bousso and J.~Polchinski, ``{Quantization of four form fluxes and dynamical
  neutralization of the cosmological constant},''
  \href{http://dx.doi.org/10.1088/1126-6708/2000/06/006}{{\em JHEP} {\bf 06}
  (2000)  006}, \href{http://arxiv.org/abs/hep-th/0004134}{{\tt
  arXiv:hep-th/0004134}}.

\bibitem{Parke:1982pm}
S.~J. Parke, ``{Gravity, the Decay of the False Vacuum and the New Inflationary
  Universe Scenario},''
  \href{http://dx.doi.org/10.1016/0370-2693(83)91376-X}{{\em Phys. Lett. B}
  {\bf 121} (1983)  313--315}.

\end{thebibliography}\endgroup

\end{document}